\documentclass[11pt,dvips,draftcls,onecolumn]{IEEEtran}

\usepackage{psfig,amsfonts,amsmath,amssymb,color,hyperref}
\hypersetup{colorlinks=true,linkcolor=black,urlcolor=dblue,pdfpagemode=None,pdfstartview=FitH}
\definecolor{gray}{cmyk}{.2,0.2,.3,.1}
\definecolor{dred}{cmyk}{0,0.9,0.4,0.3}
\definecolor{dblue}{rgb}{0,0,0.5}
\definecolor{dgreen}{rgb}{0,0.3,0}
\definecolor{dgray}{rgb}{0.3,0.3,0}
\DeclareOldFontCommand{\rm}{\normalfont\rmfamily}{\mathrm}
\DeclareOldFontCommand{\sf}{\normalfont\sffamily}{\mathsf}
\DeclareOldFontCommand{\tt}{\normalfont\ttfamily}{\mathtt}
\DeclareOldFontCommand{\bf}{\normalfont\bfseries}{\mathbf}
\DeclareOldFontCommand{\it}{\normalfont\itshape}{\mathit}
\DeclareOldFontCommand{\sl}{\normalfont\slshape}{\@nomath\sl}
\DeclareOldFontCommand{\sc}{\normalfont\scshape}{\@nomath\sc}

\newtheorem{proposition}{Proposition}

\title{On Multiflows in Random Unit-Disk Graphs, and the Capacity of
  Some Wireless Networks
  \footnote{The authors are with
  the School of Electrical and Computer Engineering, Cornell University,
  Ithaca, NY.  URL:
  \href{http://cn.ece.cornell.edu/}{{\tt http://cn.ece.cornell.edu/}}.
  Work supported by the National Science Foundation, under awards
  CCR-0238271 (CAREER), CCR-0330059, and ANR-0325556.  Parts of this
  work were presented at the 2003 edition of ACM MobiHoc, in Annapolis,
  MD~\cite{PerakiS:03}; and at the 2004 edition of the Information
  Theory Workshop, in San Antonio, TX~\cite{PerakiS:04}.}}
\author{Christina Peraki \hspace{2cm} Sergio D.\ Servetto}
\date{March 20, 2005.}

\begin{document}
\maketitle

\begin{picture}(0,0)
\put(-10,210){\tt\small Submitted to the IEEE Transactions on Information
  Theory, March 2005.}
\end{picture}

\begin{abstract}
\noindent\it
We consider the capacity problem for wireless networks.  Networks
are modeled as random unit-disk graphs, and the capacity problem is
formulated as one of finding the maximum value of a multicommodity
flow.  In this paper, we develop a proof technique based on which
we are able to obtain a tight characterization of the solution to
the linear program associated with the multiflow problem, to within
constants independent of network size.  We also use this proof method
to analyze network capacity for a variety of transmitter/receiver
architectures, for which we obtain some conclusive results.  These
results contain as a special case (and strengthen) those of Gupta
and Kumar for random networks, for which a new derivation is provided
using only elementary counting and discrete probability tools.
\end{abstract}

\section{Introduction}
\label{sec:intro}

\subsection{The Capacity of Wireless Networks -- Five Years Later}

In March 2000 (exactly five years ago as of the writing of this paper),
Gupta and Kumar published a landmark piece of work, where they presented
a thorough study on the capacity of wireless networks~\cite{GuptaK:00}.
For {\em random} networks, this problem was formulated as one of
forming tessellations of a sphere, then defining routes in between
cells, for which tight upper and lower bounds were obtained on their
capacity.  The main finding in~\cite{GuptaK:00} was actually a rather
negative one: under a variety of very reasonable scenarios, in all
cases the throughput available to each node in the network was found
to be of the form $\Theta\big(\frac{1}{\sqrt{n}}\big)$ {\em at most},
for a network with $n$ nodes -- that is, this throughput becomes
vanishingly small for large networks.

The results of~\cite{GuptaK:00} generated a flurry of activity in
this area (surveyed below, in subsection~\ref{sec:related-work}).
However, five years later, although some progress has been made
towards understanding the capacity of large networks in a regime
in which the minimum distance among nodes remains fixed and the
area covered grows unbound with the number of nodes, some questions
related to the original setup in~\cite{GuptaK:00}, dealing with
{\em high-density} networks (meaning, networks with a growing
number of nodes covering a fixed finite area) still remain, at
best, only partially answered:
\begin{itemize}
\item The ability to generate directed beams of energy in a wireless
  network could potentially change its behavior rather drastically,
  making the network ``look like'' a wired one.  What exactly
  is the impact of directional antennas then on network capacity?
\item Despite some attempts, a pure information theoretic analysis
  on the capacity of high-density wireless networks still remains
  elusive.
\end{itemize}
In this paper, we revisit the problem of capacity for random
networks considered in~\cite{GuptaK:00}.  We consider an entirely
different problem formulation: our formalization of the network
capacity problem consists of finding the value of a multicommodity
flow problem defined on a random graph, for which we are able to
obtain a number of results that contain those of~\cite{GuptaK:00}
as a special case, generalizing them in a number of interesting
directions.

\subsection{Problem Formulation}

\begin{figure}[ht]
\centerline{\psfig{file=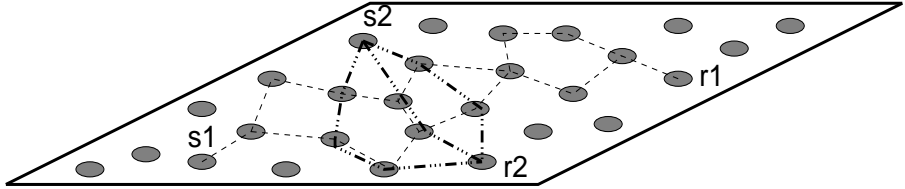,width=12cm,height=2.5cm}}
\caption{\small Problem setup.  $n$ randomly located transmitters
  send data to $n$ randomly chosen receivers, all nodes act as
  sources/destinations/relays, and nodes can only exchange messages
  with nearby nodes (within range $d_n$).}
\label{fig:problem-setup}
\end{figure}

Consider the following network communication problem.  $n$ nodes are
uniformly distributed on the closed set $[0,1]\times[0,1]$, forming
a random graph $G=(V,E)$.  Each node $s_i$ can only send messages to
and receive messages from nodes within distance $d_n$, where $d_n$,
in order for the graph to be connected with probability 1 (as
$n \rightarrow \infty$), has to satisfy
\begin{equation}
  \pi d_n^2 = \frac{\log n+\xi_n}{n}, 
  \label{eq:min-conn-radius} 
\end{equation}
for some $\xi_n\rightarrow\infty$~\cite{GuptaK:98}.  Source-destination
pairs are formed randomly: for each source node $s_i$ one destination
node $t_i$ is chosen by sampling uniformly (without replacement) from
the set of network nodes ($1 \leq i \leq n$) -- each node is both a
source, a destination for some other node, and a relay for other nodes.
All links have the same fixed finite capacity $c$, independent of network
size.  This scenario is illustrated in Fig.~\ref{fig:problem-setup}.
 
Our goal in this paper is to determine the rate of growth of the {\em
maximum stable throughput} (MST) for the network~\cite{TsybakovM:79}---the
rate at which all sources can inject packets, while maintaining stability
for the system---and provided all sources inject data at the same rate.

The problem of determining MST under a fairness constraint is an
instance of a {\em multicommodity flow} problem~\cite[Ch.\ 29]{CormenLRS:01}:
\begin{itemize}
\item There are $n$ commodities: the packets available for transmission
  from transmitter $s_i$ to receiver $t_i$.
\item The load on a single link contributed by all sources that use that
  link cannot exceed its capacity.
\item Subject to these constraints, we want to find the largest number
  of packets per unit of time that can be injected simultaneously by all
  sources.
\end{itemize}
Representing our network by a graph $G=(V,E)$, the capacity of an 
edge $e=(u,v)$ by $c(u,v)$, and letting our optimization variables be 
$f_i(u,v)$ (the flow along edge $(u,v)$ for the $i$-th commodity), then 
the maximum multiflow problem above can be formulated as a 
linear program, as shown in Table~\ref{tab:lp-multicommodity}.
 
\begin{table}[ht]
\caption{\small\rm Linear programming formulation of the multicommodity
  flow problem with a fairness constraint.}
\begin{center}
\fbox{\begin{minipage}{8.6cm}
max \hspace{2cm} $\lambda_n$ \\
$\;$ subject to:
\[\begin{array}{rll}
 & \lambda_n = \sum_{(s_i,v)\in E} f_i(s_i,v), & 1 \leq i \leq n \\
 & \sum_{i=1}^n f_i(u,v) \leq c(u,v), & (u,v) \in E \\
 & f_i(u,v) = -f_i(v,u), & (u,v)\in E, 1\leq i\leq n \\
 & \sum_{v\in V} f_i(u,v) = 0, & u\in V-\{s_i,t_i\}, 1\leq i\leq n
\end{array}\]
\end{minipage}}\end{center}
\label{tab:lp-multicommodity}
\end{table}

Our main task in this paper is to provide a characterization of the
optimal value $\lambda^*_n$.  Note that since the graph is random, and
the LP is a function of the random graph, $\lambda^*_n$ is a random
variable itself.

\subsection{Asymptotically Tight Bounds}

\begin{figure}[ht]
\centerline{\psfig{file=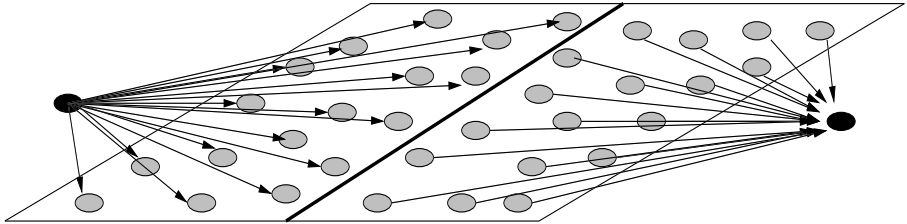,width=12cm,height=3cm}}
\caption{\small In this formulation, we only consider the traffic
  generated by sources on the left-half of the network, with destination
  on the right-half---the traffic generated by all other
  source/destination pairs is discarded.}
\label{fig:special-case}
\end{figure}

Not much is known about the structure of optimal solutions to the
maximum multiflow problem---the only technique we are aware
of for deciding whether a particular amount of flow of each commodity
can be supported by the network consists of formulating this problem
as a linear program, and then answering the non-emptyness question
for its polytope of optimization using a standard LP solver (e.g.,
the Ellipsoid method~\cite{GroetschelLS:88}), or some of the efficient
algorithms for maximum multiflow such as that of Karger and
Plotkin~\cite{KargerP:95}.  Hence, we will not be able to use those
formulations to do much more than obtain numerical results for
our problem.  We are thus motivated to search for an alternative
formulation of the problem.  And one such possible alternative is
illustrated in Fig.~\ref{fig:special-case}.

Note that doing this amounts to introducing a restriction in the domain
of optimization of the linear program from Table~\ref{tab:lp-multicommodity}:
instead of considering all possible network flows, we only consider those
which satisfy the constraints of Fig.~\ref{fig:special-case}.  But what
is crucial in this case is that, different from the problem of
Fig.~\ref{fig:problem-setup}, this new problem involving flow going
from the left to the right admits a regular {\em single commodity} flow
formulation.  The resulting linear program is shown in
Table~\ref{tab:lp-regflow}.
 
\begin{table}[ht]
\caption{\small\rm Linear program for the single commodity flow problem.} 
\begin{center}\fbox{\begin{minipage}{7cm}
max \hspace{1.5cm} $n\nu_n$ \\
subject to: 
\[\begin{array}{rll}
 & n\nu_n = \sum_{u\in V} f(s,u), & u \in V \\
 & f(u,v) \leq c(u,v), & (u,v) \in E \\
 & f(u,v) = -f(v,u), & (u,v)\in E \\  
 & \sum_{v\in V} f(u,v) = 0, & u\in V-\{s,t\} \\
\end{array}\]  
\end{minipage}}\end{center}
\label{tab:lp-regflow}
\end{table} 

The interest in this new linear program is due to the fact that,
since it corresponds to a classical single commodity problem, we
can try to solve it analytically using the max-flow/min-cut
theorem~\cite{FordF:62}.  However, the relationship between the
optimal value $\lambda^*_n$ for the ``difficult'' multicommodity
problem, and $n\nu^*_n$ for the ``easier'' single commodity problem,
is not entirely straightforward.
On one hand, the linear program in
Table~\ref{tab:lp-regflow} is a {\em restriction} of that in
Table~\ref{tab:lp-multicommodity}, since in the former some flow
variables are constrained to 0 (that is how we incorporate the
constraint of flow going only from left to right).  On the other
hand, the linear program in Table~\ref{tab:lp-regflow} is a
{\em generalization} of that in Table~\ref{tab:lp-multicommodity},
since the latter removes the multicommodity constraints (all
commodities are treated as a single commodity).  Thus, an
important question is that of giving a precise relationship between
$\lambda^*_n$ and $\nu^*_n$.
 
\subsection{Related Work}
\label{sec:related-work}

This work is primarily motivated by our struggle to understand the
results of Gupta and Kumar on the capacity of wireless
networks~\cite{GuptaK:00}.  And the main idea behind our approach
is simple: the transport capacity problem posed in~\cite{GuptaK:00},
in the context of random networks, is essentially a throughput
stability problem---the goal is to determine how much data can be
injected by each node into the network while keeping the system
stable---, and this throughput stability problem admits a very
simple formulation in term of flow networks.  Note also that because
of the mechanism for generating source/destination pairs, all
connections have the same average length (one half of one network
diameter), and thus we do not need to deal with the bit-meters/sec
metric considered in~\cite{GuptaK:00}.

As mentioned before,~\cite{GuptaK:00} sparked significant interest
in these problems.  Follow up results from the same group were reported
in~\cite{GuptaK:01, XieK:02}.  Some information theoretic bounds
for large-area networks were obtained in~\cite{LevequeT:05}.  When
nodes are allowed to move, assuming transmission delays proportional
to the mixing time of the network, the total network throughput is
$O(n)$, and therefore the network can carry a non-vanishing rate per
node~\cite{GrossglauserT:02}.  Using a linear programming formulation,
non-asymptotic versions of the results in~\cite{GuptaK:00} are given
in~\cite{ToumpisG:02}; an extended version of that work can be found
in~\cite{Toumpis:PhD}.  An alternative method for deriving transport
capacity was presented in~\cite{KulkarniV:04}.  The capacity of large
Gaussian relay networks was found in~\cite{GastparV:05}.  Preliminary
versions of our work based on network flows have appeared
in~\cite{PerakiS:03, PerakiS:04}; and network flow techniques have
been proposed to study network capacity problems (cf.,
e.g.,~\cite{AhlswedeCLY:00}, \cite[Ch.\ 14.10]{cover-thomas:it-book}),
and network coding problems~\cite{KoetterM:03}.  From the network
coding literature, of particular relevance to this work is the work
on multiple unicast sessions~\cite{LiL:04b}.

\subsection{Main Contributions and Organization of the Paper}

Let $\lambda^*_n$ denote an optimal solution to the linear program
in Table~\ref{tab:lp-multicommodity}, and let $\nu^*_n$ denote an
optimal solution to the linear program in Table~\ref{tab:lp-regflow}.
Our first result consists of finding the asymptotic value of
$\lambda^*_n$: with probability $1$ as $n\to\infty$,
\begin{equation}
  \Theta(\lambda^*_n) \;\; = \;\; \Theta(\nu^*_n)
    \;\; = \;\; \Theta\left(\frac{\log^{\frac 3 2}(n)}{\sqrt{n}}\right).
  \label{eq:main-result}
\end{equation}
This result formally establishes the equivalence between the two
linear programs, in a well defined sense: they both have solutions
which differ by at most a constant factor, independent of network
size.

A second important contribution is to show an application of the
proof methods developed to establish~(\ref{eq:main-result}), to
obtain the maximum stable throughput for various transmitter/receiver
architectures:

\begin{itemize}

\item We consider first the case of omnidirectional antennas, where
  we show that the scaling laws obtained based on our proof method
  are identical to those of~\cite{GuptaK:00}: per-node throughput is
  $\Theta\big(\frac{1}{\sqrt{n\log n}}\big)$.

\item Then we apply the same proof techniques to the determination
  of scaling laws for a new architecture, in which transmitter nodes
  can generate a single and arbitrarily narrow directed beam, and in
  which receivers can successfully decode multiple transmissions as
  long as the transmitters are not co-linear.  And in this case we
  find that:
  \begin{itemize}
  \item If only enough power to maintain the network connected is radiated
    at each node, the maximum stable throughput of this network is
    $\Theta\left(\!\sqrt{\frac{\log n}{n}}\,\right)$.
  \item If now enough power is radiated to achieve MST
    linear in network size (certainly feasible with narrow beams), then
    the number of {\em resolvable} beams that
    each node must generate is $\Theta(n)$.
  \end{itemize}

\item Finally, we consider a node architecture in which each node
  is able to generate {\em multiple and arbitrarily narrow} directed
  beams, simultaneously to all nodes within its transmission range,
  and receivers operate as above.  In this case we find that:
  \begin{itemize}
  \item If only enough power to maintain the network connected is
    radiated at each node, the maximum stable throughput of this network
    is $\Theta\big(\!\log^{\frac 3 2}n \big/ \sqrt{n}\big)$.
  \item If now enough power is radiated to achieve MST
    linear in network size (certainly feasible with 
    narrow beams), then the number of {\em resolvable} beams
    that each node must generate is $\Theta(n^{\frac 1 3})$.
  \end{itemize}

\end{itemize}

Essentially, our results show that both directional antennas, as well
as the ability to communicate simultaneously with multiple nodes, can
only achieve modest improvements in terms of achievable MST.  While
some performance gains are certainly feasible at reasonable complexities
(in the order of a low-degree polynomial in $\log n$), the number of
{\em resolvable} beams that need to be generated to increase the
achievable MST by more than a polylog factor is polynomial in network
size, and thus exponential in the minimum number of beams required to
keep the
network connected.  How many beams need to be resolved is a reasonable
measure of complexity, since the higher this number, the narrower
these beams need to be made, and hence the higher the complexity
of a practical implementation.

We also believe another original contribution is given by our proof
techniques:
\begin{itemize}
\item Our results are obtained using only elementary network flow
  concepts, and the calculations involved require only basic probability
  theory, calculus and combinatorics.  By formulating the
  problem of~\cite{GuptaK:00} as an elementary problem of flows
  in random graphs, we obtain what we believe is a number of interesting
  insights into the nature of this problem which were not obvious
  to us from their proof technique, as well as a set of meaningful
  generalizations to deal with the case of directional antennas.
\item Except for some elements of the protocol model considered
  in~\cite{GuptaK:00}, most of the work we are aware of on this
  subject (e.g.,~\cite{GastparV:05, GrossglauserT:02, GuptaK:01,
  LevequeT:05, ToumpisG:02, XieK:02}), has focused on the use
  of ``continuous'' tools, dealing with Gaussian signals, power
  constrained channels, etc.  Our work instead takes a ``discrete''
  approach to the network capacity problem, tackling it primarily
  using flow, counting and discrete probability tools.  Thus, we
  believe our proof technique, while using only elementary tools,
  has some novelty in the context of the problem considered here.
\end{itemize}
An added benefit of our proof method is that we are able to prove
{\em strong} convergence (meaning, convergence with probability 1)
in all cases.  In particular, when considering the specialization
of our results to the setup of~\cite{GuptaK:00}, our results are
stronger, in that only {\em weak} convergence is established there.

The rest of this paper is organized as follows.  In
Section~\ref{sec:bounds-formulation} we formulate upper and lower
bounds on the value of $\lambda_n^*$: the upper bound is evaluated
in Section~\ref{sec:eval-upper-bound}, and the lower bound is
evaluated in Section~\ref{sec:eval-lower-bound}.  Then,
applications of these results in the context of wireless networking
problems follow: in Section~\ref{sec:omnidirectional} we present
an alternative derivation for the results of Gupta and Kumar in
the context of random networks~\cite{GuptaK:00}, and in
Section~\ref{sec:directional} these results are extended to deal
with two different cases involving directional antennas.  The
paper concludes with Section~\ref{sec:conclusions}.

\section{Asymptotically Tight Bounds on the Value of the Linear Program}
\label{sec:bounds-formulation}

In this section we start with some preliminaries presenting the
tools used to carry out our analysis, to then go on to formulate
upper and lower bounds on the value of $\lambda_n^*$.

\subsection{Tools}
\label{sec:tools}

To compute the maximum value of a single commodity flow in our network,
we use a standard result in flow networks: the max-flow/min-cut
theorem of Ford and Fulkerson~\cite{FordF:62}.  We solve this problem
by counting how many edges can be constructed so that they all
simultaneously straddle a minimum cut.

\subsubsection{The Max-Flow/Min-Cut Theorem}

\begin{figure}[ht]
\vspace{-2mm}
\centerline{\psfig{file=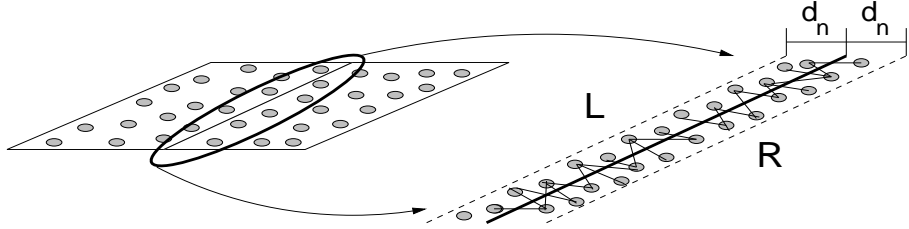,height=3cm,width=12cm}}
\vspace{-2mm}
\caption{\small To illustrate the choice of a cut to derive bounds.
  $L$ and $R$ are sections of the network on each side of the cut
  boundary, of width $d_n$, the transmission range.}
\label{fig:our-cut}
\end{figure}

$f$ is a flow of maximum value {\em iff} $|f| = c(S,T)$ (for some cut $(S,T)$).
We focus our attention on one particular cut (shown also in
Fig.~\ref{fig:our-cut}):
\begin{eqnarray*}
S & = & { (x_i,y_i)\in V \cap [0,\mbox{\small $\frac 1 2$})\times[0,1]}, \\
T & = & { (x_i,y_i)\in V \cap [\mbox{\small $\frac 1 2$},1] \times [0,1] }.
\end{eqnarray*}

In this way, to compute the value of a maximum flow we need to
determine how many edges straddle the $x=\frac 1 2$ cut.  To do
that, we proceed in two steps.  First, we compute the {\em expected}
number of edges that straddle this cut, with this expectation taken
as an ensemble average over all possible network realizations.  Then,
we derive a sharp concentration result: given an arbitrary network
realization, with probability 1 as $n\rightarrow\infty$, we show
that in this network, the actual number of edges that straddle the
cut has the exact same rate of growth (in the $\Theta$ sense
of~\cite{GrahamKP:94}) as the ensemble mean does.

\subsubsection{Mean Values}

What is the average number of nodes in a subset
$A\subseteq[0,1]\times[0,1]$?  A simple calculation shows
that
\begin{equation}
E(\textsl{Number of nodes in }A) = n P(A)
   =  n \int_A f_{XY} \mbox{d}A = n|A|,
        \label{eq:num-nodes}
\end{equation}
where $|A|$ denotes the area of $A$.

\subsubsection{Chernoff Bounds}

In addition, to prove sharp concentration results, we need to bound
the probability of deviations from its mean by sums of independent
random variables:
\begin{itemize}
\item Consider $n$ points $X_{1} \ldots X_{n}$ iid and uniformly distributed
  on $[0,1] \times [0,1]$.  We have a number of
  subsets $A_j \subset [0,1] \times [0,1]$, for $j=1...f(n)$ (the number
  of subsets may depend on the number of points $n$), and denote the area
  of any such subset by $|A_j|$.  Now we define some random variables:
  \[ N_{ij} = \left\{ \begin{array}{ll}
               1, & X_i \in A_j \\
               0, & \textrm{otherwise.}
              \end{array} \right.
  \]
  Since the $X_i$'s are independent, the $N_{ij}$'s are also independent.
\item Now let $N_j$ be another random variable defining the number of points
  in $A_j$, i.e., $N_j=\sum_{i=1}^n N_{ij}$.  We see in this case that the
  $N_j$'s, $j=1 \ldots f(n)$ are random variables where each is the sum of
  $n$ iid binary random variables (but not necessarily independent among
  the $N_j$'s themselves).
\item The expected number of points in $A_{j}$ is
  \[ E(N_j) \;\; = \;\; E\left(\sum_{i=1}^n N_{ij}\right)
            \;\; = \;\; \sum_{i=1}^n E(N_{ij}). \]
  But, since $P(X_i \in A_j) = |A_j|$, we have that
  $E(N_{ij}) = 1|A_j|+0(1-|A_j|) = |A_j|$, and hence
  $E(N_j) = n|A_j|$.
\end{itemize}

For the family of variables $N_j$, we have the following standard
results, known as the {\em Chernoff} bounds (see, e.g.,~\cite[Ch.\
4]{MotwaniR:95}):
\begin{enumerate}
\item For any $\delta >0$:
\[ P\big[N_j>(1+\delta)n|A_j|\big]
   \;\; < \;\; \left(\frac{e^{\delta}}{(1+\delta)^{1+\delta}}\right)^{n|A_j|}.
\]
\item For any $0<\delta <1$:
\[ P\big[N_j<(1-\delta)n|A_j|\big] \;\; < \;\; e^{-\frac{1}{2}n|A_j|\delta^{2}}.
\]
\end{enumerate}
With a few simple calculations we can rewrite the first bound as
\begin{eqnarray*}
\lefteqn{P\big[(N_j-n|A_j|)>\delta n|A_j|\big]
  \;\; < \;\;
    \left(\frac{e^{\delta}}{e^{(1+\delta)\ln(1+\delta)}}\right)^{n|A_j|}} \\
  & = & \left(e^{\delta - (1+\delta)\ln(1+\delta)}\right)^{n|A_j|}
  \;\; = \;\; e^{(\delta - (1+\delta)\ln(1+\delta))n|A_j|}
  \;\; = \;\; e^{-\theta_{1} n|A_j|},
\end{eqnarray*}
where $-\theta_{1} \triangleq \delta-(1+\delta)\ln(1+\delta)$.  We can
also rewrite the second bound as
\[ P\big[(N_j - n|A_j|)<-\delta n|A_j|\big]
     \;\; < \;\; e^{(-\frac{1}{2}\delta^{2})n|A_j|}
     \;\; = \;\; e^{ -\theta_{2} n|A_j|},
\]
where $-\theta_2 \triangleq -\frac{1}{2} \delta^{2}$.  Consider
now the case of $0<\delta<1$: restricted to this range, we have
that $\theta_1 > 0$; and $\theta_2$ is clearly positive as well.
Thus, by defining $\theta(\delta) = \min(\theta_1,\theta_2)$,
we have
\begin{equation}
  P\big[\;|N_j-n|A_j|| > \delta n|A_j|\;\big] \;\; < \;\; e^{-\theta n|A_j|}.
  \label{eq:chernoff}
\end{equation}
Our interest in~(\ref{eq:chernoff}) is because, if we can prove
probability bounds of that form, then we can claim that
$\frac{N_j}n = \Theta(|A_j|)$ with probability 1, in the limit
as $n\to\infty$.  In other words, for the random variables $N_j$,
as $n\to\infty$, there exist constants such that deviations from
their mean by more than these constants occur with probability 0.
Note that as $n\to\infty$, $e^{-\theta n|A_j|} \rightarrow 0$, so
\[\lim_{n\to\infty} P\big[\;|N_j-n|A_j|| > \delta n|A_j|\;\big] \;\;=\;\; 0, \]
or equivalently,
\begin{eqnarray*}
1 & = & \lim_{n\to\infty} P\big[\;|N_j-n|A_j|| \leq \delta n|A_j|\;\big] \\
  & = & \lim_{n\to\infty} P\big[\;0 \leq (1-\delta)n|A_j| \leq N_j
          \leq (1+\delta) n|A_j|\;\big] \\
  & = & \lim_{n\to\infty} P\big[N_j=\Theta(n|A_j|)\;\big],
\end{eqnarray*}
where $\Theta$ is defined in~\cite{GrahamKP:94} as:
\[
\Theta(g(n)) \;\; = \;\;
  \big\{ f(n) : \exists c_1>0, c_2>0, n_0, \textrm{ for which } 0 \leq
  c_1 g(n) \leq f(n) \leq c_2 g(n) \textrm{,  } \forall n \geq n_0. \big\}.
\]

\subsection{An Equivalent Linear Program}
\label{sec:ssd}

As suggested in the Introduction, we will not work directly with
the original linear program from Table~\ref{tab:lp-multicommodity},
but instead we will work with a new linear program, one in which
flow is constrained to move from left to right.  The new LP is
formally stated in Table~\ref{tab:demand-zero}.

\begin{table}[!ht]
\caption{\small\rm Linear programming formulation of the multicommodity
  flow problem, in which the supply of connections other than those going
  from left to right are set to 0.}
\begin{center}\fbox{\begin{minipage}{12cm}
max \hspace{2cm} $\ell_n$ \\
$\;$ subject to:
\[\begin{array}{rll}
 & \ell_n = \sum_{(s_i,v)\in E} f_i(s_i,v), & 1\leq i\leq n \textrm{ and }\\
               && s_i \in S=\{u \in V: u \in [0,\frac{1}{2})\times[0,1]\},\\
               && t_i \in T=\{u \in V: u \in [\frac{1}{2},1]\times[0,1]\},\\
 & \sum_{i=1}^n f_i(u,v) \leq c(u,v), & (u,v) \in E \\
 & f_i(u,v) = -f_i(v,u), & (u,v)\in E, 1\leq i\leq n \\
 & \sum_{v\in V} f_i(u,v) = 0, & u\in V-\{s_i,t_i\}, 1\leq i\leq n \\
 & f_i(s_i,v) = 0,
   & \forall s_i\in T=\{u\in V:u\in[\frac{1}{2},1]\times[0,1]\},\\
   && t_i\in S=\{u\in V:u\in[0,\frac{1}{2})\times[0,1]\}.
\end{array}\]
\end{minipage}}\end{center}
\label{tab:demand-zero}
\end{table}

Considering sources on the left half of the network and destinations
on the right, essentially says that in our linear programming formulation
in Table~\ref{tab:lp-multicommodity} we must add the constraint of
setting to 0 the demands of commodities such that either the source
is located on the right or the sink is located on the left.  But this
constraint changes the result of the linear program only by a constant
factor, and therefore asymptotically we get the same values from
Table~\ref{tab:lp-multicommodity} and Table~\ref{tab:demand-zero}.
Intuitively, the reason is simple: since nodes are uniformly distributed,
we should have about $n/2$ nodes in $S$ and about $n/2$ nodes in $T$
with high probability; at the same time, since the source/destination
pairs are uniformly distributed, about $n/4$ of the sources are placed
on the left side of the network with destinations on the right side;
therefore, by considering only traffic generated by sources in $S$
for destinations only in $T$ the value of the multicommodity problem
should at most decrease by a factor of 4, and hence remains of the
same order.  To see this more formally, consider the following indicator
variables:
\[ I^{(n)}_i = \left\{ \begin{array}{ll}
               1, & s_i \in S \wedge t_i \in T \\
               0, & \textrm{otherwise,}
              \end{array} \right.
  \]
where $S$ and $T$ are given in Table~\ref{tab:demand-zero}.  Then,
$I^{(n)}=\sum_{i=1}^{n} I^{(n)}_i$ is another random variable whose
value is equal to the number of pairs with the source on the left half
and the sink on the right half.  We would like to compute how many
are these pairs, to calculate the difference between the values of
the two linear programs.  To do this, we first compute the mean of
$I^{(n)}$, then we use the Chernoff bounds to prove a sharp concentration
of this variable around its mean.

We start by computing $E(I^{(n)})$.  We have:
 \[ E(I^{(n)}) \;\; = \;\; E\left(\sum_{i=1}^n I^{(n)}_i\right)
         \;\; = \;\; \sum_{i=1}^n E(I^{(n)}_i), \]
due to linearity of expectation.  Now,
 \[ E(I^{(n)}_i)
    \;\; = \;\; 1 \cdot P(s_i \in S \wedge t_i \in T)
                + 0 \cdot P(s_i \in T \vee t_i \in V)
    \;\; = \;\; P(s_i \in S \wedge t_i \in T). \]
Since the nodes in our network are uniformly and independently
distributed, we have that the events $\{s_i\in S\}$ and $\{t_i\in T\}$
are independent events, and therefore:
\[ E(I^{(n)}_i) \;\; = \;\; P(s_i \in S \wedge t_i \in T)
          \;\; = \;\; P(s_i \in S) \cdot P(t_i \in T)
          \;\; = \;\; \frac{1}{2} \times \frac{1}{2}
          \;\; = \;\; \frac{1}{4},\]
which finally gives us:
\[ E(I^{(n)}) \;\; = \;\; \sum_{i=1}^n E(I^{(n)}_i)
        \;\; = \;\; \frac{n}{4}.\]
This expected number occurs with high probability as $n\to\infty$ because,
according to the Chernoff bound,
\[
 P\big[\;|I^{(n)}-E(I^{(n)})| > \delta E(I^{(n)})\;\big]
  \;\; < \;\; e^{-\theta E(I^{(n)})}
  \;\; = \;\; e^{-\theta \frac{n}{4}} \;\; \to \;\; 0,
\]
as $n \rightarrow \infty$.  Thus, with high probability, there
are about $\frac{n}{4}$ sources in $S$ with destinations in $T$.
As a result, from the fairness constraint we have that
$P(\ell_n^* = \lambda_n^*/4)\to 1$, and herefore,
$\Theta(\lambda_n^*) = \Theta(\ell_n^*)$.

\subsection{Formulation of the Bounds}

With these tools, it is easy to describe asymptotic upper and lower bounds
on $\lambda_n^*$:
\begin{itemize}
\item To obtain an upper bound, we eliminate the multicommodity
  constraints from the linear program in Table~\ref{tab:demand-zero},
  and use the max-flow/min-cut theorem to compute the value of a maximum
  flow.  Thus, we have that $\Theta(\ell_n^*) \leq \Theta(\nu_n^*)$,
  for $\nu_n$ as defined in Table~\ref{tab:lp-regflow}.
\item To obtain a lower bound, we construct a feasible point for the
  linear program in Table~\ref{tab:demand-zero}, to obtain a value
  $\gamma_n$ for the LP for which, clearly, $\gamma_n \leq \ell_n^*$.
\end{itemize}
In the next two sections we evaluate these bounds, to show that
\[
  \Theta\left(\frac{\log^{\frac 3 2}n}{\sqrt{n}}\right)
  \;\; = \;\; \gamma_n
  \;\; \leq \;\; \ell_n^*
  \;\; \leq \;\; \Theta(\nu_n^*)
  \;\; = \;\; \Theta\left(\frac{\log^{\frac 3 2}n}{\sqrt{n}}\right),
\]
and thus conclude that
$\ell_n^* = \Theta\left(\frac{\log^{\frac 3 2}n}{\sqrt{n}}\right)$,
and thus
$\lambda_n^* = \Theta\left(\frac{\log^{\frac 3 2}n}{\sqrt{n}}\right)$
as well.

\section{Evaluation of the Upper Bound}
\label{sec:eval-upper-bound}

In this section, our goal is to show that $\nu_n^* = 
\Theta\left(\frac{\log^{\frac 3 2}n}{\sqrt{n}}\right)$, based on
the methods outlined in the previous section.

\subsection{Counting Edges Across a Minimum Cut}
\label{sec:upperb-average}

\begin{figure}[ht]
\centerline{\psfig{file=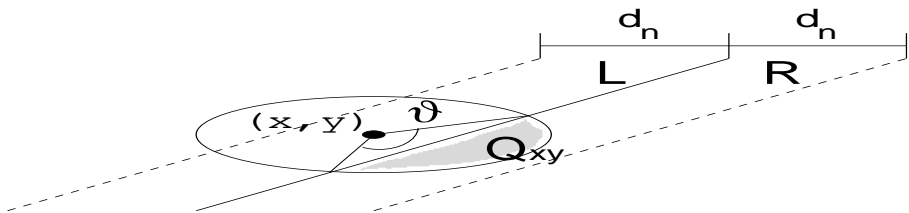,width=12cm,height=2.7cm}}
\caption{\small To illustrate constraints on edges.}
\label{fig:multiple-beams-count}
\end{figure}

Fix a particular node on the left side of the minimum cut, $L$.
The number of edges that cross the cut for that one node is exactly
the number of nodes in the right side of the cut, $R$, that are
within distance $d_{n}$. Therefore, for an arbitrary point $p=(x,y)$
in $L=[\frac 1 2-d_n,\frac 1 2)\times [0,1]$, we draw a circle of
radius $d_n$ and center $(x,y)$.  The points $q=(u,v)$ in $R=[\frac
1 2,\frac 1 2+d_n]\times [0,1]$ that are inside the circle are equal
to the number of edges we want to count.  These points $p$ and $q$
for which an edge exists satisfy the following conditions: (1)
$\frac 1 2-d_n \leq x \leq \frac 1 2$; (2) either (a) $0 \leq y \leq
1$, or (b) $d_n \leq y \leq 1-d_n$; (3) $\frac 1 2 < u$; and (4)
$(u-x)^2 + (v-y)^2 \leq d_n^2$.
The situation is illustrated in Fig.~\ref{fig:multiple-beams-count}.\\

For each $p=(x,y)$, we get the average number of points $q=(u,v)$
within the shaded arc $Q_p$ in Fig.~\ref{fig:multiple-beams-count}
using eqn.~(\ref{eq:num-nodes}): $E(${\sl Number of points in} $Q_p)
= n|Q_p|$.

To compute the area of $Q_p$ (denoted $|Q_p|$), we let $\vartheta$
denote the angle of the arc illustrated in
Fig.~\ref{fig:multiple-beams-count}. Then, it follows from
elementary trigonometric identities that
$\sin{\frac{\pi-\vartheta}{2}}=\frac{\frac{1}{2}-x}{d_{n}}$, and so
$\cos{\frac{\vartheta}{2}}=\frac{\frac{1}{2}-x}{d_{n}}$.  So, $|Q_p|
  = \mbox{\small$\frac 1 2$} \vartheta d_{n}^2
    -\mbox{\small$\frac 1 2$}
    d_n\cos{\frac{\vartheta}{2}}2d_{n}\sin{\frac{\vartheta}{2}}
  = \mbox{\small$\frac 1 2$} \vartheta d_n^2
    - \mbox{\small$\frac 1 2$}d_n^2\sin{\vartheta}
  = \mbox{\small$\frac 1 2$}d_n^2 (\vartheta - \sin{\vartheta})$.
And plugging this expression into $n|Q_p|$, we get $n|Q_p| = n
\frac{1}{2} d_n^2 (\vartheta - \sin{\vartheta})$. But from the
trigonometric identities above, we have that $\vartheta = 2
\arccos{\frac{\frac{1}{2}-x}{d_{n}}}$ and hence, $\sin{\vartheta} =
2 \sin{\frac{\vartheta}{2}} \cos{\frac{\vartheta}{2}}$, which
implies $\sin^2{\vartheta} =
4\sin^2{\frac{\vartheta}2}\cos^2{\frac{\vartheta}2}$, which again
implies $\sin^2{\vartheta}
  = 4 (1 -\cos^2{\frac{\vartheta}{2}}) \cos^2{\frac{\vartheta}{2}}$.
Now, since $0 \leq \vartheta \leq \pi$, $\sin{\vartheta}$ $=$ $2
\cos{\frac{\vartheta}{2}}
                   \sqrt{1 -\cos^2{\frac{\vartheta}{2}}}$ $\geq$ $0$,
and so, finally, we get an expression for $n|Q_p|$ in terms of $n$,
$d_n$, and the coordinates of the transmitter $p=(x,y)$:
\begin{eqnarray*}
n|Q_p|
  & = & \frac{1}{2} n d_n^2 (\vartheta - \sin{\vartheta}) \\
  & = & \frac{1}{2} n d_n^2 \left(2 \arccos{\frac{\frac{1}{2}-x}{d_{n}}} -
        2\cos{\frac{\vartheta}{2}}\sqrt{1-\cos^2{\frac{\vartheta}{2}}}\right)\\
  & = & n d_{n}^2 \left(\arccos{\frac{\frac{1}{2}-x}{d_{n}}} -
                      \frac{\frac{1}{2}-x}{d_{n}}
                      \sqrt{1 -\frac{(\frac{1}{2}-x)^2}{d_{n}^2}}\right).
\end{eqnarray*}

The result above is the average number of edges that cross
the cut, starting at a fixed point $p=(x,y)$ in $L$.  To calculate
the total number of edges $S$ that cross the cut on average, we need
to add up $n|Q_p|$ over all nodes $p$, (i.e., compute $S =
\sum_{p\in L} n|Q_p|$).
And our plan to do this is to approximate this sum by an integral.

The value of $|Q_p|$ is clearly dependent on the location of $p$:
for $p$'s in $L$ near the boundary of the cut ($x<\approx\frac 1
2$), $\vartheta\approx\pi$ and hence the shaded area is large; for
$p$'s still in $L$ but far from the boundary of the cut
($x>\approx\frac 1 2-d_n$), $\vartheta\approx 0$ and hence the
shaded area is small.  Furthermore, except near the top and bottom
boundaries, the area of $Q_p$ is independent of $y$.  Therefore, to
obtain a simple expression for the sought sum, our first step
consists of dividing $L$ into $\frac{d_n}\Delta$ thin strips of
height 1 and width $\Delta$ (for $\Delta\ll d_n$), and expanding
$\sum_{p\in L} n|Q_p|$ in two different ways:
\begin{eqnarray*}
S_a & = & \sum_{k=1}^{d_n/\Delta}  n|Q_{xy}| \cdot\underbrace{|\{
p=(x,y)\in L:
     0\leq y\leq 1 \}|}_{s_a}; \\
S_b & = & \sum_{k=1}^{d_n/\Delta}  n|Q_{xy}| \cdot\underbrace{|\{
p=(x,y)\in L:
     d_n\leq y\leq 1-d_n \}|}_{s_b};
\end{eqnarray*}
(in both cases, we take $\frac 1 2-d_n+(k-1)\Delta\leq x\leq \frac 1
2-d_n+k\Delta$).  $S_a$ is an upper bound on $\sum_{p\in L} n|Q_p|$,
since we may count edges that end up outside the network; $S_b$ is a
lower bound, since we may not count some valid edges close to the
network boundary; but as long as $d_n \rightarrow 0$ as
$n\rightarrow\infty$, both bounds become tight
and equal to $\sum_{p\in L} n|Q_p|$.

The next step is to observe that once again we can approximate the
size estimates $s_a$ and $s_b$ using eqn.~(\ref{eq:num-nodes}): $s_a
= n\Delta$ and $s_b = n(1-2d_n)\Delta$.  Hence we get:
\begin{eqnarray*}
S_a & = & \sum_{k=1}^{d_n/\Delta}  n|Q_{xy}|\cdot n\Delta
    \;\;\; = \;\;\; n^2\Delta\sum_{k=1}^{d_n/\Delta} |Q_{xy}| \\
    & \approx & n^2\int_{x=\frac 1 2-d_n}^{\frac 1 2}\int_{y=0}^1
                |Q_{xy}|\mbox{d}x\mbox{d}y; \\
S_b & = & \sum_{k=1}^{d_n/\Delta}  n|Q_{xy}|\cdot n(1-2d_n)\Delta
    \;\;\; = \;\;\; n^2(1-2d_n)\Delta\sum_{k=1}^{d_n/\Delta} |Q_{xy}| \\
    & \approx & n^2\int_{x=\frac 1 2-d_n}^{\frac 1 2}\int_{y=d_n}^{1-d_n}
                |Q_{xy}|\mbox{d}x\mbox{d}y,
\end{eqnarray*}
since $\Delta\sum_{k=1}^{d_n/\Delta} |Q_{xy}|$ is a Riemann sum
that, as we let $\Delta\rightarrow 0$, converges to the integral
over an
appropriate region of $|Q_p|$.

And now we are almost done.  Since $S_b \leq \sum_{p\in L} n|Q_p|
\leq S_a$, and we have that for $n$ large, $S_a \approx S_b \approx
n^2\int_L |Q_p|$, we finally get:
\begin{eqnarray*}
\lefteqn{\sum_{p\in L} n|Q_p| \;\;\approx\;\; n^2\int_L |Q_p|\mbox{ d}p} \\
  & = &  n^2 \int_{\frac{1}{2}-d_{n}}^{\frac{1}{2}}
    \int_{0}^{1} d_{n}^2 \left\lbrack \arccos{\frac{\frac{1}{2}-x}{d_{n}}}
    -
    \frac{\frac{1}{2}-x}{d_{n}} \sqrt{1 -\frac{(\frac{1}{2}-x)^2}{d_{n}^2}}
    \right\rbrack \mbox{d}y \mbox{d}x \\
  & = & n^2 d_{n}^2 \int_{\frac{1}{2}-d_{n}}^{\frac{1}{2}} \int_{0}^{1}
    \arccos{\frac{\frac{1}{2}-x}{d_{n}}}\mbox{ d}y \mbox{d}x
    - n^2 d_{n}^2
    \int_{\frac{1}{2}-d_{n}}^{\frac{1}{2}}
    \int_{0}^{1} \frac{\frac{1}{2}-x}{d_{n}}
    \sqrt{1 -\frac{(\frac{1}{2}-x)^2}{d_{n}^2}}\mbox{ d}y \mbox{d}x  \\
  & = & n^2 d_{n}^2 \int_{\frac{1}{2}-d_{n}}^{\frac{1}{2}}
    \arccos{\frac{\frac{1}{2}-x}{d_{n}}} \mbox{ d}x
    - n^2 d_{n}^2
    \int_{\frac{1}{2}-d_{n}}^{\frac{1}{2}} \frac{\frac{1}{2}-x}{d_{n}}
    \sqrt{1 -\frac{(\frac{1}{2}-x)^2}{d_{n}^2}} \mbox{ d}x \\
  & \stackrel{(a)}{=} & -n^2 d_{n}^3 \int_{1}^{0} \arccos{u} \mbox{ d}u +
    n^2 \int_{1}^{0} u \sqrt{ 1 - u^2} \mbox{ d}u \\
  & = & n^2 d_{n}^3 \int_{0}^{1} \arccos{u} \mbox{ d}u -
        n^2 \int_{0}^{1} u \sqrt{1 - u^2} \mbox{ d}u \\
  & = & n^2 d_{n}^3  - \mbox{\small $\frac 1 3$}n^2d_n^3
  \;\; = \;\; \mbox{\small $\frac 2 3$}n^2d_n^3,
\end{eqnarray*}
where $(a)$ follows from the change of variable
$\frac{\frac{1}{2}-x}{d_{n}}=u$.

\subsection{Sharp Concentration Results}

Our next goal is to show that the actual number of edges
straddling the cut in any realization of the network is sharply
concentrated around its mean.  That is, in almost all networks, the
number of edges across the cut is $\Theta(n^2d_n^3) =
\Theta(\sqrt{n} \log^\frac{3}{2}(n)).$

Define a binary random variable $N_{ij}$, which takes the value 1
if the $i$-th node is within the transmission range of a node at
coordinates $(x_j,y_j)$ on the other side of the cut, as illustrated
in Fig.~\ref{fig:multiple-beams-count}:
\[ N_{ij} = \left\{ \begin{array}{rl}
                    1, & X_i \in Q_{(x_j,y_j)} \\
                    0, & \textrm{otherwise.}
                    \end{array}
            \right.
\]
Let $p$ denote the probability that $X_i$ is in $Q_{(x_j,y_j)}$
(i.e., that $N_{ij} = 1$).  Then, $p = |Q_{(x_j,y_j)}| = \frac 1 2
d_n^2(\vartheta-\sin(\vartheta))$, with $0\leq\vartheta\leq\pi$ is
as in Fig.~\ref{fig:multiple-beams-count}.  Therefore, defining
$\kappa_\vartheta$ as $\frac 1 2 (\vartheta-\sin(\vartheta))$, we
have $p = |Q_{(x_j,y_j)}| = \kappa_\vartheta d_n^2 =
\kappa_\vartheta\frac{\log n}n$.

Define $N_j = \sum_{i=1}^n N_{ij}$ as the number of points in
$Q_{(x_j,y_j)}$.  In this case, we have $E(N_j)$ = $E\big(\sum_{i=1}^n
N_{ij}\big)$ = $\sum_{i=1}^n p\cdot 1 + (1-p)\cdot 0$ = $np$ =
$\kappa_\vartheta\log(n)$. Now, by eqn.~(\ref{eq:chernoff}), we have
that
\[
  P\big(|N_j-\kappa_\vartheta\log(n)| > \delta\kappa_\vartheta\log(n)\big)
  \;\; < \;\; e^{-\theta\kappa_\vartheta\log(n)}
  \;\; = \;\; n^{-\theta\kappa_\vartheta},
\]
As $n\rightarrow\infty$ this probability tends to zero, and
therefore, in almost all network realizations, a node on the left
side of the cut is connected to $\Theta(\log(n))$ nodes on the
right side.\footnote{Observe that $\kappa_\vartheta=0$ only over
a set of measure zero (the set of network locations such that
$x=\frac 1 2 - d_n$), and thus the exponent can be assumed
strictly positive.}
By an analogous argument, we have $\Theta(nd_n) =
\Theta\big(\sqrt{n\log n}\big)$ nodes on the left half.
Therefore, the actual number of edges across
the cut is $\Theta(\log^2 n)\cdot\Theta\big(\sqrt{n\log n}\big)$,
so $n\nu_n^* = \Theta\big(\sqrt{n}\log^{\frac 3 2} n\big)$.

\section{Evaluation of the Lower Bound}
\label{sec:eval-lower-bound}

To give a lower bound for $\ell_n^*$, we construct one feasible
point: this is accomplished by giving a specific routing algorithm,
and finding how much traffic this scheme can carry:
\begin{enumerate}
\item We start by proving that, with probability 1 as $n\to\infty$,
  there is a subgraph of the random graphs under consideration with
  a clear, regular structure.
\item We then develop a (very simple) routing technique that makes
  use of the links in the regular subgraph only.
\item Finally, we determine the throughput achieved in this way.
\end{enumerate}

\subsection{Existence of a Regular Subgraph}

\begin{figure}[!ht]
\centerline{\psfig{file=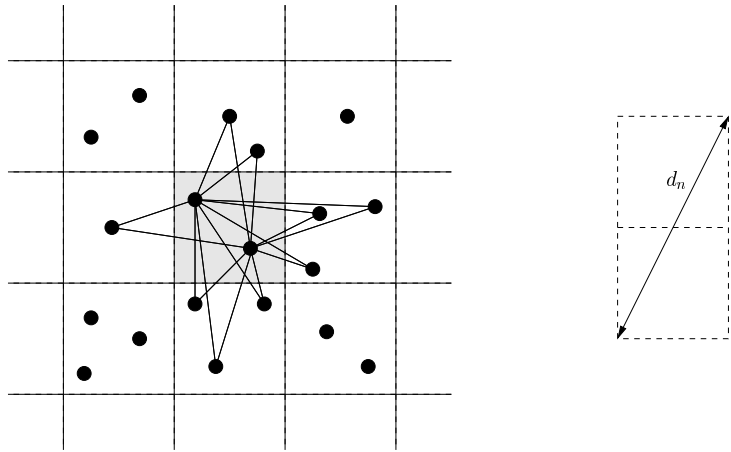,height=6cm}}
\caption{\small To illustrate the presence of a structured subgraph
  for large $n$.  Consider the shaded center cell: all nodes within
  that cell are connected by an edge to every node in the cells above,
  below, left and right.  To guarantee that all such edges can be
  formed, from Pithagoras, the transmission range $d_n$ must be
  chosen as $d_n = \sqrt{\frac{5c\log n}{n}}$.  So, provided $c>0$,
  connectivity of the network is guaranteed~\cite{GuptaK:98}.}
\label{fig:regular-subgraph}
\end{figure}

Consider a partition of the network area (the closed set
$[0,1]\times[0,1]$) into {\em square} cells, each one of area
$c\frac{\log n}{n}$.  To determine $c$, we observe that
the side of a cell is $\sqrt{\frac{c\log n}{n}}$, and so $c$
is chosen such that the inverse of this number,
$\sqrt{\frac{n}{c\log n}}$, is an integer -- this is to
guarantee that the cells form a partition of the whole
network, as illustrated in Fig.~\ref{fig:regular-subgraph}.

With the construction of grid and choice of connectivity radius
$d_n$ shown in Fig.~\ref{fig:regular-subgraph}, each node within
a cell will have an edge connecting it to all nodes in the four
adjacent cells.

\subsection{A Routing Algorithm}

\begin{figure}[!ht]
\centerline{\psfig{file=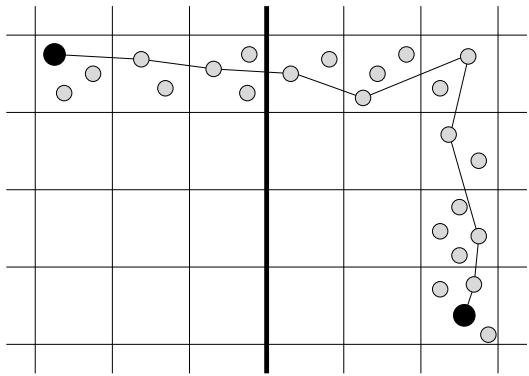,height=5cm}}
\caption{\small To illustrate routing mechanics.  A source in a cell
  left of the center cut sends messages to a destination on the right
  side by first forwarding data horizontally, then vertically.  Under
  the assumption that all cells are non-empty, there is always a next
  hop.}
\label{fig:routing-mechanics}
\end{figure}

To describe the routing algorithm, we define first some notation:
\begin{itemize}
\item Given a cell $(i,j)$
  ($1\leq i,j\leq \sqrt{\frac 1 c \frac{n}{\log n}}$), $v_{ij}$ denotes
  any arbitrary node $v\in V$ contained in that cell.
\item Given a node $v\in V$, $(i(v),j(v))$ denotes the cell that contains $v$.
\item $v^{\textrm{\tiny curr}}$: current node; $v^{\textrm{\tiny dest}}$:
  destination node.
\end{itemize}
The algorithm executed at each node to decide the next hop of a message
is as follows:
\begin{enumerate}
\item If $j(v^{\textrm{\tiny curr}}) < j(v^{\textrm{\tiny dest}})$,
    send message to $(i(v^{\textrm{\tiny curr}}),j(v^{\textrm{\tiny curr}})+1)$.
\item Else, if $i(v^{\textrm{\tiny curr}}) > i(v^{\textrm{\tiny dest}})$,
    send message to $(i(v^{\textrm{\tiny curr}})-1,j(v^{\textrm{\tiny curr}}))$.
\item Else, if $i(v^{\textrm{\tiny curr}}) < i(v^{\textrm{\tiny dest}})$,
    send message to $(i(v^{\textrm{\tiny curr}})+1,j(v^{\textrm{\tiny curr}}))$.
\item Else, $v^{\textrm{\tiny curr}}$ and $v^{\textrm{\tiny dest}}$ are in
  the same cell (so $v^{\textrm{\tiny dest}}$ is reachable in one hop from
  $v^{\textrm{\tiny curr}}$), hence stop.
\end{enumerate}
These mechanics are illustrated in Fig.~\ref{fig:routing-mechanics}.

For this algorithm to work properly, we must insure that no cells
are empty: if this condition holds, then we can be sure that an L-shaped
path as shown in Fig.~\ref{fig:routing-mechanics} will always deliver
packets to destination.  But the fact that no cells are empty is not
obvious, and requires proof.  In fact, we will prove something stronger:
the number of nodes contained in any arbitrary cell is $\Theta(\log n)$,
with high probability as $n\to\infty$.

Define $X_{ij}$ as the number of nodes within a cell 
$(i,j)$ ($1\leq i,j\leq \sqrt{\frac 1 c \frac{n}{\log n}}$).  Then,
\begin{itemize}
\item Mean value of $X_{ij}$:
  \begin{equation}
  E(X_{ij}) = n \cdot \frac{c\log n}{n} = c\log n.
  \label{mean_sensors_per_node}
  \end{equation}
\item Chernoff bound on deviations from the mean for $X_{ij}$:
  \[
  P(|X_{ij}-c\log n| > \delta c\log n)
    \;\; \leq \;\; e^{-\theta c\log n}
    \;\; = \;\; \frac 1{n^{c\theta}}.
  \]
\item Probability that the occupancy of {\em none} of the cells 
  deviates significantly from its mean:
  \begin{eqnarray*}
  P\left(\bigcap_{i,j} |X_{ij}-c\log n| < \delta c\log n\right)
     & = & 1 - P\left(\bigcup_{i,j}
            |X_{ij}-c\log n| > \delta c\log n\right) \\
     & \geq & 1 - \sum_{i,j}
            P\big(|X_{ij}-c\log n| > \delta c\log n\big).
  \end{eqnarray*}
  Consider now any $\epsilon>0$; there is a value $n_0(\epsilon)$ such that,
  for all $n>n_0(\epsilon)$,
  \[
     \sum_{i,j} P\big(|X_{ij}-c\log n| > \delta c\log n\big)
     \;\; < \;\; \sum_{i,j} \frac 1{n^{c\theta}}
     \;\; \stackrel{(a)}{=} \;\; \frac 1 c \frac 1{n^{c\theta-1}\log n}
     \;\; < \;\; \epsilon,
  \]
  where $(a)$ follows from the fact that there are
  $\frac 1 c \frac n{\log n}$ cells, and provided $c > \frac 1 \theta$.
  Therefore,
  \begin{eqnarray*} 
  P\left(\bigcap_{i,j} |X_{ij}-c\log n| < \delta c\log n\right)
    & \geq & 1-\epsilon,
  \end{eqnarray*}
  and thus all cells contain $\Theta(\log n)$ nodes almost surely, as
  $n\to\infty$.
\end{itemize} 
With this, we see that all cells {\em simultaneously} will be non-empty
in almost all networks.  Thus, the routes defined by the proposed routing
algorithm will always deliver data to destination.  We still need to
determine how much though.

\subsection{Computation of the Achievable Throughput}

The last step is to determine the throughput available to a
source/destination pair constructed by the routing algorithm above.

We start by stating the relatively straightforward fact that, since
the routes constructed by the algorithm above do not split the flow
at any intermediate node, the throughput of a connection is determined
by the capacity available to that connection at the link with highest
load.\footnote{This intuitive fact can be formalized based on Robacker's
decomposition theorem for multiflows~\cite{PerakiS:04, Robacker:56}.}
Now, since links have a fixed finite capacity, and since in our problem
we work under a {\em fairness} constraint that forces all source/destination
pairs to inject the same amount of data, we have that the capacity allocated
to a commodity on any link is $\frac {\textrm{\tiny raw link capacity}}
{\textrm{\tiny \# of commodities using that link}}$.  Thus, our problem
reduces to finding the maximum number of commodities sharing a link.

We claim that no link in the network is shared by more connections
than the links which straddle the center cut:
\begin{itemize}
\item The number of commodities sharing a link across the center cut is
  exactly equal to the number of nodes within a horizontal strip, as
  illustrated in Fig.~\ref{fig:horizontal-links}.
  \begin{figure}[!ht]
  \centerline{\psfig{file=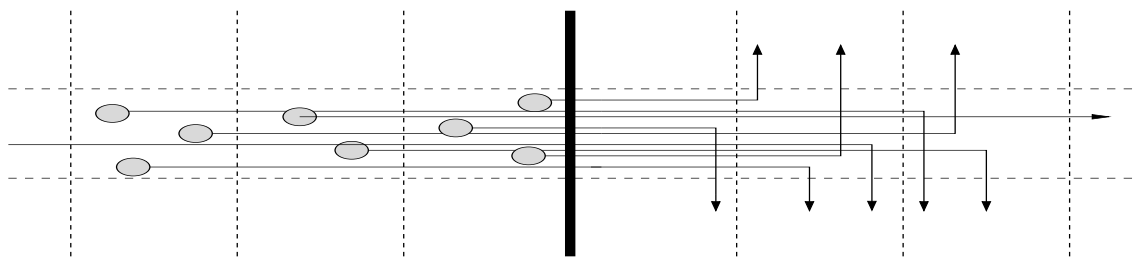,width=15cm,height=3.4cm}}
  \caption{\small Each node left of the center cut generates traffic that
    must cross that cut.}
  \label{fig:horizontal-links}
  \end{figure}
\item Clearly, the number of commodities on horizontal links (meaning,
  links going from one cell to another cell either left or right) decreases
  as we move away from the center cut:
  \begin{itemize}
  \item Moving left, the number of sources decreases.
  \item Moving right, once a connection reaches the column on which the
    cell containing its destination lies, it starts moving along vertical
    links and never goes back to horizontal ones.
  \end{itemize}
\item The number of commodities on vertical links is at most the same
  as the number on links in the center cut -- but this requires proof.
\end{itemize}
To prove this last point, we need to count the number of commodities
sharing a horizontal link crossing the center cut, and we have to give
an upper bound on the number of commodities sharing an arbitrary vertical
link.

In terms of the number of commodities sharing a horizontal link across
the center cut:
\begin{itemize}
\item The average number of commodities across the center cut is just
  \[
    n \cdot \mbox{$\frac 1 2$}\sqrt{\frac{c\log n}{n}}
    \;\; = \;\; \mbox{$\frac{\sqrt{c}}2$}\sqrt{n\log n},
  \]
  and again from the Chernoff bounds, we have that this is not only the
  ensemble average, but that in almost all networks, this number is
  $\Theta\big(\!\sqrt{n\log n}\big)$.
\item By a similar argument, we have that the number of edges across
  the center cut in between two adjacent cells is $\Theta(\log^2 n)$ --
  with high probability, $\Theta(\log n)$ nodes in each cell, by construction
  there is a link between any two of those.
\end{itemize}
Thus, the number of commodities sharing a link across the center cut is
$\frac{\Theta(\sqrt{n\log n})}{\Theta(\log^2 n)}$
= $\Theta\left(\frac{\sqrt{n}}{\log^{\frac 3 2}(n)}\right)$.

\begin{figure}[!ht]
\centerline{\psfig{file=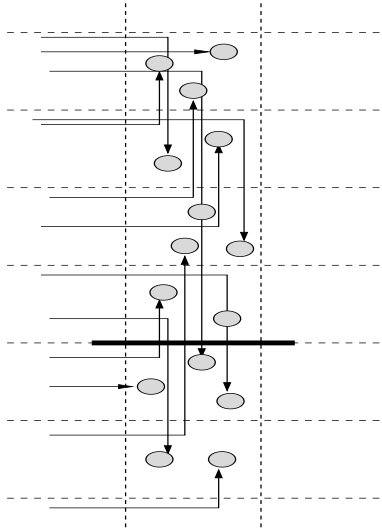,height=7cm,width=5cm}}
\caption{\small Upper bound on the number of commodities sharing a vertical
  link.  Clearly, not all commodities that use a vertical link share a link
  across the thick horizontal line: some will switch from horizontal to
  vertical above the line and reach their destination before crossing that
  line, and the same will happen below.  By estimating the number of
  commodities sharing a vertical link by the number of nodes in a vertical
  strip, we effectively say that all commodities in that strip use {\em all}
  vertical edges.  This is certainly an overestimate, based on which we
  obtain only an upper bound.}
\label{fig:vertical-links}
\end{figure}

In terms of the number of commodities sharing any vertical link,
an upper bound on this number is given by
$\frac{\textrm{\tiny \# of nodes in a vertical strip}}
      {\textrm{\tiny \# of links between adjacent cells}}$.
Why this is an upper bound is illustrated in Fig.~\ref{fig:vertical-links}.
But then, since the number of nodes in a vertical strip is twice the
number of nodes in $\frac 1 2$ of a horizontal strip, from an argument
entirely analogous to the count of commodities sharing a link across
a center cut in the paragraph above, we have that the number of commodities
sharing a vertical link is {\em at most}
$\Theta\left(\frac{\sqrt{n}}{\log^{\frac 3 2}(n)}\right)$.

In summary, we have that the link sharing the largest number of
commodities is shared by
$\Theta\left(\frac{\sqrt{n}}{\log^{\frac 3 2}(n)}\right)$ of them.
Therefore, by the fairness constraint, the capacity of this link
is shared equally among all commodities, and thus this capacity
is the sought $\gamma_n$ value, i.e.,
$\gamma_n = \Theta\left(\frac{\log^{\frac 3 2}(n)}{\sqrt{n}}\right)$ is
achievable for the linear program in Table~\ref{tab:demand-zero}.

\subsection{Remark}

Note: the ``spirit'' of this proof is very similar to the proof
in~\cite[Sec.\ IV]{GuptaK:00}: in both cases, the goal is to give
an explicit construction to show the achievability of certain
throughput values.  However, the methods employed to analyze the
throughput achieved by the routing strategies proposed differ
significantly --~\cite{GuptaK:00} relies heavily on VC
theory~\cite{VapnikC:71}, whereas we only use properties of flows
and the Chernoff bound.

\section{Applications to Wireless Networking Problems I: the Gupta-Kumar Setup}
\label{sec:omnidirectional}

Before considering more general node architectures in
Section~\ref{sec:directional}, we show in this section how, for the case
of nodes equipped with omnidirectional antennas, using our proof techniques
we obtain scaling laws identical to those reported in~\cite{GuptaK:00},
but under strong convergence.

\subsection{Transmitter/Receiver Model}

\begin{figure}[ht]
\centerline{\psfig{file=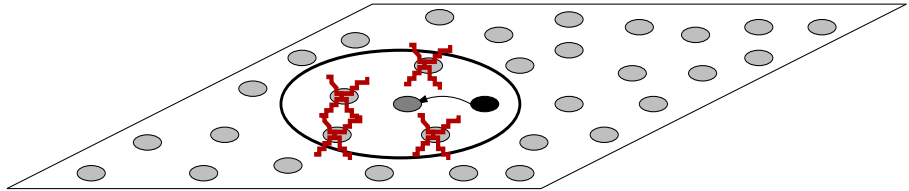,width=12cm,height=2.5cm}}
\caption{\small A transmission model based on omnidirectional antennas
  and pure collisions.}
\label{fig:txrx-model3}
\end{figure}

In~\cite{GuptaK:00}, transmissions were omnidirectional, and described
based on a pure collision model: for a transmission to be successfully
decoded, no other transmission has to be in progress within the range
of the receiver under consideration.  This setup is illustrated in
Fig.~\ref{fig:txrx-model3}.

\subsection{Average Number of Edges Across the Cut}

Our first task is to determine the {\em average} number of edges that
can be simultaneously supported across the cut, average taken over all
possible network realizations.

\subsubsection{An Upper Bound}

\begin{figure}[ht]
\centerline{\psfig{file=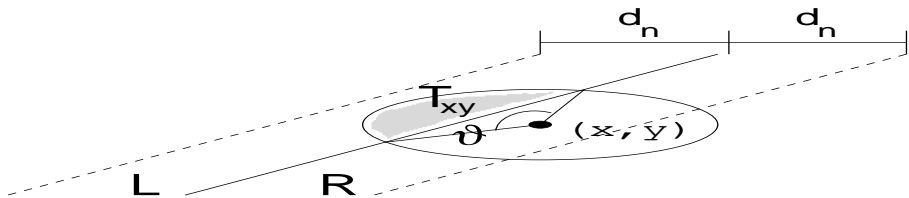,width=12cm,height=2.5cm}}
\caption{\small For a receiver at location $(x,y)$, at most one transmitter
  in the shaded region $T_{xy}$ can send a message (if this message is to be
  successfully decoded on the other side of the cut).}
\label{fig:reverse-count}
\end{figure}

For a fixed receiver location $(x,y)$ in $R$, there can only be one
active transmitter within distance $d_n$ of the receiver, for that
transmission to be successfully received.  Since to obtain an upper
bound we only need worry about edges that cross the cut, we first
consider all possible locations of one such transmitter in $L$, by
drawing a circle of radius $d_n$ and center $(x,y)$.  This region
is illustrated in Fig.~\ref{fig:reverse-count}.

Denoting by $|T_{xy}|$ the area of the shaded region $T_{xy}$ in
Fig.~\ref{fig:reverse-count}, we use eqn.~(\ref{eq:num-nodes}) to
estimate the number of transmitters located in $T_{xy}$ as $n|T_{xy}|$.
However, since only one transmitter located within $T_{xy}$ can transmit
successfully to a receiver at $(x,y)$, the number of nodes that are
able to transmit at the same time from $L$ to $R$ is upper bounded by
\[
   \frac{E(\textsl{Number of nodes in $L$})}
        {E(\textsl{Number of nodes in $T_{xy}$})}
   = \frac{n L}{n T_{xy}}  
\]
This is an {\em upper bound}, because we are assuming that it is possible
to find a set of locations $(x,y)$ in $R$ such that no area in $L$ is
wasted---showing that this bound is indeed tight requires proof.

Now, the area of $L$ is $d_n$.  To compute the area of $T_{xy}$, we
have to determine the area of an arc of a circle with angle $\vartheta$, as
shown in Fig.~\ref{fig:reverse-count}, and in a computation entirely
analogous to that of the calculation of $|Q_p|$ in
Section~\ref{sec:multiple-directed-beams}.  In this case, we have that
$\sin(\frac 1 2(\pi-\vartheta))=\frac{x-\frac 1 2}{d_n}=\cos(\frac 1 2\vartheta)$,
and since $\frac{1}{2} \leq x < \frac{1}{2}+d_{n}$ it is clear that
we must have $0 < \vartheta \leq \pi$ and also $\sin{\vartheta} \geq 0$.  Then,
we get $|T_{xy}|
    = \mbox{\small$\frac 1 2$} \vartheta d_n^2
      - \mbox{\small$\frac 1 2$} d_n \cos(\mbox{\small$\frac 1 2$}\vartheta)
      2 d_n \sin(\mbox{\small$\frac 1 2$}\vartheta)   
    = \frac{1}{2} \vartheta d_{n}^2 - \frac{1}{2} d_{n}^2 \sin{\vartheta}$,
and therefore, $|T_{xy}|
= \mbox{\small$\frac 1 2$} d_n^2 (\vartheta - \sin{\vartheta})$.  Hence,
for each possible value of $\vartheta$, an upper bound on the number of
nodes that are able to transmit at the same time from $L$ to $R$ is
\[
   \frac{nL}{nT_{xy}}=\frac{nd_n}{n\frac 1 2d_n^2(\vartheta-\sin\vartheta)}
   =\frac 2{d_n(\vartheta-\sin\vartheta)}.
\]
Since this upper bound depends on the choice of receiver location
(through the angle $\vartheta$), we will make this bound as small as possible
by an appropriate choice of $\vartheta$.  As noted above, $0<\vartheta\leq\pi$,
and $\sin\vartheta\geq 0$.  Hence, the number of transmitters in $L$ is
smallest when $\vartheta=\pi$ and $\sin\vartheta=0$, i.e., when the receivers
are located close to the cut boundary (as it should be, since it is in this
case when receivers ``consume'' the maximum amount of transmitter area).
In this case, we get
\[ \min_{0<\vartheta\leq\pi}\left[\frac{2}{d_n(\vartheta-\sin\vartheta)}\right]
   = \frac 2{\pi d_n}
\]
as an upper bound on the number of edges across the cut.  Furthermore,
in this case we see immediately that to maximize capacity we must keep
$d_n$ as small as possible---and we know from eqn.~(\ref{eq:min-conn-radius})
that the smallest possible $d_n$ that will still maintain the network
connected is $\Theta(\sqrt{\log n/n})$.  Therefore, replacing for the
optimal $d_n$, we finally get an upper bound of
$\Theta\left(\sqrt{n/\log n}\right)$.

\subsubsection{The Upper Bound is Asymptotically Tight}

\begin{figure}[ht]
\centerline{\psfig{file=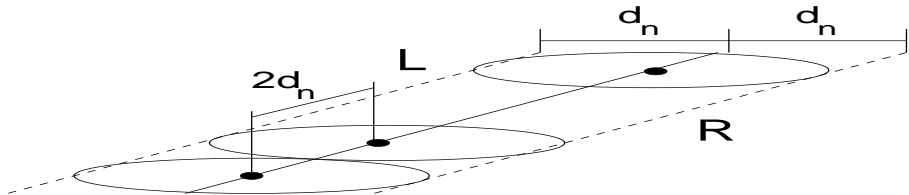,width=12cm,height=2.5cm}}
\caption{\small An explicit flow construction.}
\label{fig:explicit-flow-construction}
\end{figure}

To verify that the upper bound is tight, we give an explicit flow
construction.  Consider the placement of disks shown in
Fig.~\ref{fig:explicit-flow-construction}.

Since the height of the square is $1$, and we are placing nodes at
distance $2d_n$ from each other, this guarantees that {\em if there
are nodes in each of the circles to create valid tx/rx pairs}, then
the number of successful simultaneous transmissions across the cut
is $\frac 1{2d_n}=\Theta\left(\sqrt{n/\log n}\right)$.  Whether all such pairs
of nodes can be created simultaneously or not is the issue addressed
next.

\subsection{Uniform Convergence Issues}

Next we prove that when $n$ points are dropped uniformly over the
square $[0,1] \times [0,1]$, we have that simultaneously (i.e.,
uniformly) over all $\frac 1{2d_n}$ circles from
Fig.~\ref{fig:explicit-flow-construction}, each one of the circles
contains $\Theta(\log(n))$ points in almost all network realizations.
From this, we conclude that the distribution of the number of edges
across the cut is sharply concentrated around its mean, and hence
that in a randomly chosen network, with probability approaching 1 as
$n\rightarrow\infty$, the actual number of straddling edges is indeed
$\Theta\left(\sqrt{n/\log(n)}\right)$.

\subsubsection{Statement of the Result}

Consider we have $\frac{1}{2d_{n}}$ circles centered along the
$x=\frac 1 2$ cut as shown in Fig.~\ref{fig:explicit-flow-construction},
with centers $y_{j}=(2j-1)d_{n}$, $j=1 \ldots \frac{1}{2d_{n}}$ and radius
$d_{n}$.  Then, we have the following uniform convergence result:

\begin{proposition}
Define $B_j := [|N_{j}- \pi \log n| < \delta \pi \log n]$.
Then, as $n \rightarrow \infty$, and for any $\delta \in (x,1)$
($x \approx 0.6$), we have that
\[ \lim_{n\rightarrow\infty}
   P\left[\bigcap_{j=1}^{\sqrt{\frac{n}{\log n}}} B_j\right] = 1.
\]
\label{prop:uniformity}
\end{proposition}
Essentially what this proposition says is that with very high probability
and uniformly over $j$, all $A_j$'s contain $\Theta(\log n)$ nodes.

\subsubsection{Proof}

Note that the area of a circle in Fig.~\ref{fig:explicit-flow-construction}
is $\pi d_n^2 = \pi \frac{\log n}{n}$.  Then, from the Chernoff bound,
we have that for any $0<\delta<1$ we can find a $\theta > 0$ such that
\begin{equation}
   P[|N_j-\pi \log n| > \delta \pi \log n]
     < e^{-\theta \pi \log n} = n^{-\theta\pi}.
   \label{eq:ch1}
\end{equation}
Thus, we can conclude that the probability that the number of nodes in
a circle deviates by more than a constant factor from the mean tends to
zero as $n\rightarrow\infty$.  This is a key step in showing that all
the events $B_j := \big[|N_j-\pi\log(n)| < \delta\pi\log(n)\big]$ occur
{\em simultaneously}.  Now, from the union bound, we have that
\[
  P \left[ \bigcap_{j=1}^{\frac{1}{2d_{n}}} B_{j} \right]
    \;\; = \;\; 1 - P \left [\bigcup_{j=1}^{\frac{1}{2d_{n}}} B_{j}^{c} \right]
    \;\; \geq \;\; 1 - \sum_{j=1}^{\frac{1}{2d_{n}}} P[B_{j}^{c}].
\]
But, from eqn.~(\ref{eq:ch1}), $P[B_j^c] < n^{-\theta\pi}$, and therefore,
\[
  \sum_{j=1}^{\frac 1{2d_n}} P[B_j^c]
  \;\; < \;\; \sum_{j=1}^{\frac{1}{2d_{n}}}  n^{-\theta\pi}
  \;\; = \;\; \frac{n^{-\theta\pi}}{2d_n} 
  \;\; = \;\; \frac{n^{\frac{1}{2} - \pi \theta}}{2 \sqrt{\log n}}
\]
Putting everything together, and letting $n\rightarrow\infty$, we have
\[
   P \left[\bigcap_{j=1}^{\frac{1}{2d_n}}B_j\!\right]
   \;\;\geq\;\; 1\!-\!\frac{n^{\frac{1}{2} - \pi \theta}}{2 \sqrt{\log n}}
   \;\;\longrightarrow\;\; 1,
\]
if and only if $\pi \theta > \frac{1}{2}$.  And this is true for
$\delta \approx 0.6$ and above (this follows from the definition of $\theta$
and a simple numerical evaluation).

\section{Applications to Wireless Networking Problems II: Directional Antennas}
\label{sec:directional}

\subsection{On Directional Antennas and MST Issues}
 
We consider now an application of the techniques that were used
so far to analyze the network capacity problem in the context of
directional antennas.

Why the interest in directional antennas?  Because there is a
question about wireless networks equipped with such antennas which
we believe is very important, and for which we could not find a
satisfactory answer in the literature.  We discussed in
Section~\ref{sec:omnidirectional} the vanishing throughput problem
identified in~\cite{GuptaK:00}.  But in a different segment of the
research community, the use of {\em directional} antennas has also
received a fair amount of attention in recent times.  The rationale
is that with omnidirectional antennas, existing MAC protocols require
all nodes in the vicinity of a transmission to remain silent.  With
directional antennas however, it should be possible to achieve higher
overall throughput, by means of a higher degree of spatial reuse of
the shared medium, and a smaller number of hops visited by a packet
on its way to destination (see, e.g.,~\cite{ChoudhuryYRV:02}).
Furthermore, in the context of energy-efficient broadcast/multicast,
it has been argued that the ability of a transmitter to reach multiple
receivers is an important source of gains to take advantage of in
the development of suitable protocols, such as BIP~\cite{WieselthierNE:02}.

If we take a step back, careful reading of these previous results
raises an important question: how much exactly is there to gain from the
use of directional antennas?  Could directional antennas (in which the
width of the beams tends to zero as $n$ gets large) be used to effectively
overcome the vanishing maximum throughput of~\cite{GuptaK:00}?  Although
we have not been able to find answers to this question in the literature
(and that motivated us to start working on this problem in the first
place), we have found a couple of related results based on which we can
say a-priori that the answer is probably {\em no}:
\begin{itemize}
\item In~\cite{GuptaK:00}, the authors claim that their result holds
  irrespective of whether transmissions are omnidirectional or directed,
  provided that in the case of directed antennas there is some lower
  bound (independent of network size) on how narrow the beams can be made.
\item In~\cite{MergenT:02,TongZM:01}, for some {\em regular} networks,
  it is shown that enabling nodes with Multi-Packet Reception (MPR)
  capabilities~\cite{GhezVS:89} can only increase the total
  throughput of the network by a constant factor ($\approx 1.6$),
  independent of network size.
\end{itemize}
Given this state of affairs, it seems to us that deciding exactly how
much there is to be gained by using directional antennas, and giving
some measure of how complex the transmitters/receivers need to be made
to achieve those gains, is indeed a topic worth being studied.
 
\subsection{A Single Directed Beam}
\label{sec:single-directed-beam}

\subsubsection{Transmitter/Receiver Model}

\begin{figure}[ht]
\centerline{\psfig{file=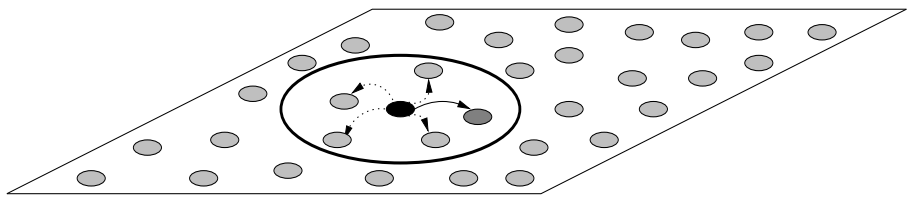,width=12cm,height=2.5cm}}
\caption{\small A single beam model for communication between nodes.}
\label{fig:txrx-model2}
\end{figure}

In this section we consider the first model based on directional
antennas: transmitters can generate a beam of arbitrarily narrow width
aimed at any particular receiver, and receivers can accept any number
of incoming messages, provided the transmitters are not in the same
straight line.  This results in a significant increase in the complexity
of the signal processing algorithms required at each node, and in this
section our goal is to determine if and how much it is possible to
increase the achievable MST, compared to the omnidirectional case.
This model is illustrated in Fig.~\ref{fig:txrx-model2}.
  
Our goal in this subsection is to evaluate $\Theta(\nu_n^*)$ and
$\Theta(\gamma_n)$, for this particular architecture.

\subsubsection{Average Number of Edges Across the Cut}

\begin{figure}[ht]
\centerline{\psfig{file=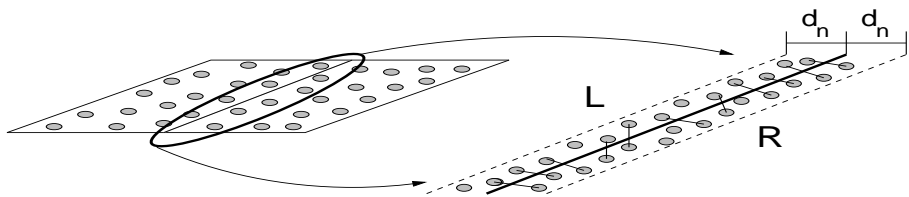,width=12cm,height=2.5cm}}
\caption{\small Pairing up one transmitter in $L$ with one receiver in
  $R$: at most $n|L| = n|R| = nd_n$ such pairs can be formed.}
\label{fig:cut-singlebeam}
\end{figure}

Since at most one edge per transmitter can be active at any point
in time, the average number of edges going across the cut can be no
larger than $n d_n$, the average number of transmitters on its left
side.  Since $L$ and $R$ have the same area, the average number of
nodes on each side of the cut is the same (and equal to $nd_n$), and
hence the maximum of $nd_n$ transmissions can actually be received,
by ``pairing up'' every node from one side of the cut with every
node on the other side.  The pairing of nodes on each side of the
cut is illustrated in Fig.~\ref{fig:cut-singlebeam}.

Finally we note that, under the assumption of arbitrarily narrow and
perfectly aligned beams, the only way in which we could have multiple
receivers blocked out by a single transmission is by having them all
lying in a nearly straight line (i.e., a set of vanishing measure)
under the beam of a single transmitter.  But then, to have an actual
edge count lower than $\Theta(nd_n)$, we would require an increasingly
large number of nodes falling in a decreasingly small area: under our
statistical model for node placement, this event occurs with vanishing
probability, and therefore the average edge count is $\Theta(nd_n)$.

\subsubsection{Sharp Concentration Results}

\paragraph{Number of Transmitters in $L$ and Receivers in $R$}
Again, consider $n$ points $X_1...X_n$ uniformly distributed over
the $[0,1]\times [0,1]$ plane, and consider the area $L$ on the left
side of the cut, as shown in Fig.~\ref{fig:cut-singlebeam}.  We define
variables
\[ N_i = \left\{\begin{array}{rl}
         1, & X_i \in L \\
         0, & \textrm{otherwise.}
         \end{array}\right.
\]
and $N = \sum_{i=1}^n N_i$.  The probability $p$ of $X_i\in L$ is
$p = |L| = 1\cdot d_n$.  Hence,
$E(N_i) = 1\cdot p + 0\cdot(1-p) = p = d_n$,
and $E(N) = \sum_{i=1}^n E(N_i) = nd_n$.
From the Chernoff bound, we know that
\[ P\left(|N-nd_n| > \delta nd_n\right) \;\; < \;\; e^{-\theta nd_n}.
\]
Since $\theta>0$, we have that as $n\rightarrow\infty$, deviations of
$N$ from its mean by a constant fraction (independent of $n$) occur
with low probability, provided $d_n$ does not decay too fast.\footnote{Note
that the fastest possible decay for $d_n$, according to
eq.~(\ref{eq:min-conn-radius}), is when $d_n \approx \sqrt{\frac{c\log n}n}$.
And in this case, $e^{-\theta nd_n} = e^{-\theta\sqrt{cn\log n}}\to 0$
as $n\to\infty$.  If $d_n$ is any bigger, this probability goes to zero
even faster.  So the Chernoff bound applies for any connected network.}
Therefore, we conclude that in almost all realizations of the network,
the number of transmitters in $L$ and the number of receivers in $R$ is
$\Theta(nd_n)$.

\begin{figure}[ht]
\centerline{\psfig{file=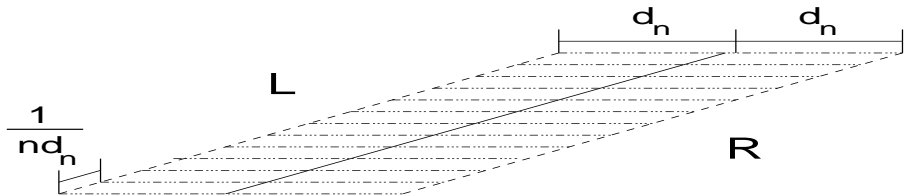,width=12cm,height=2.5cm}}
\caption{\small Cutting $L$ and $R$ into rectangles of size $d_n \times
  \frac 1{nd_n}$.}
\label{fig:cutting-L}
\end{figure}

\paragraph{Number of Edges Across the Cut}
Knowing that we have $\Theta(nd_n)$ transmitters and receivers within
range of each other on each side of the cut is not enough to claim that
the number of edges that cross the cut is $\Theta(nd_n)$.  This is because,
in our model for directional antennas, a receiver can successfully decode
two simultaneous incoming transmissions provided the angle formed by the
receiver and the two transmitters is strictly positive: if all three are
on the same straight line, collisions still occur, and those edges are
destroyed.  Therefore, we still need to show that the actual number of
edges is $\Theta(nd_n)$.  And to do this, we need to say something about
the location of points that end up in $L$, and not just count how many.
To proceed, we cut the area of $L$ into $nd_n$ rectangles of height
$\frac 1{nd_n}$ and width $d_n$, as illustrated in Fig.~\ref{fig:cutting-L}.
Our goal then becomes to show that in ``most'' of these rectangles (meaning,
in all but a constant fraction of them) we will have nodes capable of
forming straddling edges.

Counting how many of the $nd_n$ rectangles in Fig.~\ref{fig:cutting-L}
contain at least one of the $\Theta(nd_n)$ nodes that are dropped in $L$
is an instance of a classical {\em occupancy} problem, in which $k$ balls
are thrown uniformly onto $m$ bins, in the case where
$k=m=nd_n$~\cite[Ch.\ 4]{MotwaniR:95}.  Since $\frac 1 m$ is the
probability that a ball falls in any particular bin, the probability $p$
of an empty bin after throwing all $m$ balls is $p=(1-\frac 1 m)^m$ which,
for $m$ large, becomes approximately $\frac 1 e$.  Therefore, the average
number of empty bins is $mp \approx \frac 1 e \sqrt{n\log(n)}$.  And by
the Chernoff bound, again we have that
\[ P\left(Y- nd_n/e > \delta nd_n/e\right)
   < e^{-\theta nd_n/e},
\]
where $Y$ is the number of empty bins.  So, the probability that the
number of empty bins is a constant factor away from its mean is small
(again, provided $d_n$ does not decay too fast), and hence, for $n$
large, almost all network realizations will have $\Theta(nd_n)$ non-empty
rectangles.  But since transmitter/receiver pairs in different rectangles
are not collinear, the number of edges across the cut is $\Theta(nd_n)$,
qed.

\subsubsection{Remarks}

\paragraph{MST in a Minimally Connected Network}
In this section, we found that the MST achievable by the type of
tx/rx pairs considered here depends on the connectivity radius $d_n$.
If we replace $d_n$ with $\sqrt{\frac{c\log n}{n}}$ (the minimum radius
required to maintain a connected network, from~\cite{GuptaK:98}), we
get
\[ n d_n \approx n \sqrt{\frac{c\log n}{n}}
         = \Theta\left(\!\sqrt{n \log n}\right).
\]
Comparing this expression with its equivalent from
Section~\ref{sec:omnidirectional}, we see that all we gain over the
case of omnidirectional antennas is an increase in MST by a
factor of $\Theta(\log n)$.

\paragraph{Minimum Connectivity Radius Resuting in MST = $\Theta(n)$}
In this tx/rx architecture we are considering the use of arbitrarily
narrow and perfectly aligned directed beams.  Therefore, it does make
sense to consider the use of a possibly larger transmission range than
the minimum required to keep the network connected, since in this case
a large range does not force other tx/rx pairs to remain
silent while a given transmission is in progress.  And since by increasing
the transmission range now we can increase throughput, our next goal
is to determine the minimum range that would be required to achieve
MST = $\Theta(n)$.

Solving for $d_n$ in $\Theta(n) = \Theta(nd_n)$, we see that
trivially, $d_n = \Theta(1)$.  That is, to achieve MST linear in the
number of nodes using a single beam in each transmission, the radius
of each transmission has to be a constant independent of $n$.

\paragraph{Minimum Number of Simultaneous Beams}
From a practical point of view, does it matter that to achieve linear
MST we need to keep the transmission radius constant?  In this section
we argue that yes it does, very much.  To see why this is so, next we
count the minimum number $\beta$ of narrow beams that a transmitter
would have to generate simultaneously, if MST linear in the size of
the network is to be achieved: this number gives a measure of the
{\em complexity} of the beamforming transmitter, since $2\pi/\beta$ is
an upper bound on the maximum angle of dispersion of the beam.

Since a node can generate a beam to any receiver within its transmission
range (see Fig.~\ref{fig:txrx-model2}), again using eqns.~(\ref{eq:num-nodes})
and~(\ref{eq:chernoff}), we have that for $n$ large, the number of points
within a circle of radius $d_n$ is $\Theta(n\cdot\pi d_n^2)$.  In the case
of $d_n$ only satisfying the requirement of keeping the network connected,
\[
  \beta \;\; = \;\; n\cdot\pi d_n^2
        \;\; = \;\; n\left(\frac{\pi c\log n}n\right)
        \;\; = \;\; \Theta(\log n).
\]
This fact was known already---see~\cite{XueK:02} for a more complete
analysis (constants hidden by the $\Theta$-notation included), including
also a number of interesting references on the history of this problem.
But if now we consider a larger $d_n$ satisfying the requirement of
achieving linear MST, then
\[
  \beta \;\; = \;\; n\cdot\Theta(1)^2 \;\; = \;\; \Theta(n).\]
Therefore, we see $\beta$ has an {\em exponential} increase relative
to the number required to maintain minimum connectivity---it is on this
fact that we base our claim about directional antennas not being able
to provide an effective means of overcoming the issue with per-node
vanishing throughputs.

\subsection{Multiple Directed Beams}
\label{sec:multiple-directed-beams}

\subsubsection{Transmitter/Receiver Model}

\begin{figure}[ht]
\centerline{\psfig{file=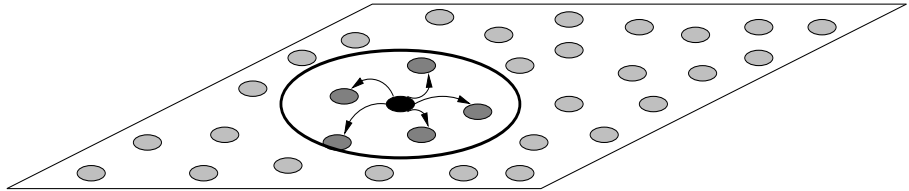,width=12cm,height=2.5cm}}
\caption{\small A non-degraded broadcast channel model for communication
  between nodes: each node is able to send simultaneously a different packet
  to each one of the nodes within his transmission range.  Furthermore,
  multiple broadcasts (from different transmitters) do not collide, unless
  the transmitters are perfectly aligned.}
\label{fig:txrx-model1}
\end{figure}

In this section we consider another model based on directional
antennas: transmitters can generate an arbitrary number of beams, of
arbitrarily narrow width, aimed at any particular receiver; and receivers
can accept any number of incoming messages, provided the transmitters are
not in the same straight line.  This is perhaps the most complex scheme
that could be envisioned based on directed beams.  Our goal is to determine
if and how much it is possible to increase the achievable MST, compared
to the previous two cases.  This model is illustrated in
Fig.~\ref{fig:txrx-model1}.

Again, our goal in this subsection is to evaluate $\Theta(\nu_n^*)$ and
$\Theta(\gamma_n)$, for this particular architecture.

\subsubsection{Average Number of Edges Across the Cut}

With minor variations, this calculation is essentially identical
to that presented in Section~\ref{sec:upperb-average}, and the
final result is the same: the ensemble average number of edges
straddling the center cut is $\Theta(n^2d_n^3)$.  See
Section~\ref{sec:upperb-average} for details.

\subsubsection{Sharp Concentration Results}

Our next goal is to show that the actual number of edges straddling
the cut in any realization of the network is sharply concentrated around
its mean.  That is, in almost all networks, the number of edges across
the cut is $\Theta(n^2d_n^3)$,

\paragraph{Number of Receivers per Transmitter}
Define a binary random variable $N_{ij}$, which takes the value 1 if
the $i$-th node is within the transmission range of a node at coordinates
$(x_j,y_j)$ on the other side of the cut, as illustrated in
Fig.~\ref{fig:multiple-beams-count}:
\[ N_{ij} = \left\{ \begin{array}{rl}
                    1, & X_i \in Q_{(x_j,y_j)} \\
                    0, & \textrm{otherwise.}
                    \end{array}
            \right.
\]
Let $p$ denote the probability that $X_i$ is in $Q_{(x_j,y_j)}$ (i.e.,
that $N_{ij} = 1$).  Then, $p = |Q_{(x_j,y_j)}| =
\frac 1 2 d_n^2(\vartheta-\sin(\vartheta))$, with $0\leq\vartheta\leq\pi$
is as in Fig.~\ref{fig:multiple-beams-count}.  Therefore, defining
$\kappa_\vartheta$ as $\frac 1 2 (\vartheta-\sin(\vartheta))$, we have
$p = |Q_{(x_j,y_j)}| = \kappa_\vartheta d_n^2 =
\kappa_\vartheta\frac{\log n}n$.

Define $N_j = \sum_{i=1}^n N_{ij}$ as the number of points in
$Q_{(x_j,y_j)}$.  In this case, we have $E(N_j) = \sum_{i=1}^n N_{ij}
= \sum_{i=1}^n p\cdot 1 + (1-p)\cdot 0 = np = \kappa_\vartheta\log(n)$.
Now, again from the Chernoff bound, we have that
\[
  P(|N_j-\kappa_\vartheta\log(n)| > \delta\kappa_\vartheta\log(n))
  < e^{-\theta\kappa_\vartheta\log(n)} = n^{-\theta\kappa_\vartheta},
\]
for $\theta$ defined as in previous applications.  As $n\rightarrow\infty$
this probability tends to zero, and therefore, in almost all network
realizations, a transmitter on the left side of the cut will be able to
reach $\Theta(\log(n))$ receivers on the right side.

\begin{figure}[ht]
\centerline{\psfig{file=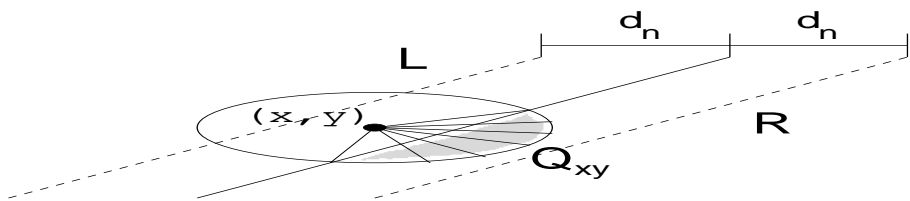,width=12cm,height=2.5cm}}
\vspace{-1mm}
\caption{\small Cutting the shaded arc $Q_{xy}$ into regions of area
  $\frac 1 n$, to formulate this as an occupancy problem analogous to
  that of Fig.~\ref{fig:cutting-L}.}
\label{fig:cutting-Qp}
\end{figure}

\paragraph{Total Number of Edges}
In a manner analogous to the situation discussed in
Section~\ref{sec:single-directed-beam}, knowing that there are
$\Theta(nd_n)$ transmitters on the left side of the cut, and that each
transmitter can reach $\Theta(nd_n^2)$ receivers on the other side, is
not enough to conclude that the total number of edges going across the
cut must be $\Theta(n^2d_n^3)$.  This is because of our requirement that
multiple transmitters not be perfectly aligned with a receiver for this
receiver to decode all these messages simultaneously.  Therefore, we
still need to show that the actual number of edges is $\Theta(n^2d_n^3)$.
And to do this, we need to say something about the location of points in
$R$ that can be reached from $L$, and not just count how many.  To
proceed then, we cut the area of $Q_p$ into $\kappa_\vartheta\log(n)$
slices, each slice of area $\frac{|Q_p|}{\kappa_\vartheta\log(n)}
= \frac 1 n$, as illustrated in Fig.~\ref{fig:cutting-Qp}.

As in the occupancy problem considered in
Section~\ref{sec:single-directed-beam}, our goal is to show that in
``most'' of these arc slices (most meaning, in all but a constant fraction
of them) we will have nodes capable of forming straddling edges.  This
is again a problem of throwing $k$ balls uniformly into $m$ bins, where
$k = m = \kappa_\vartheta\log(n)$.  And again, we have that with probability
that tends to 1 as $n\rightarrow\infty$, the number of empty bins is
$\kappa_\vartheta\log(n)/e$, and hence the number of occupied bins is
$\Theta(\log(n))$.

\begin{figure}[ht]
\centerline{\psfig{file=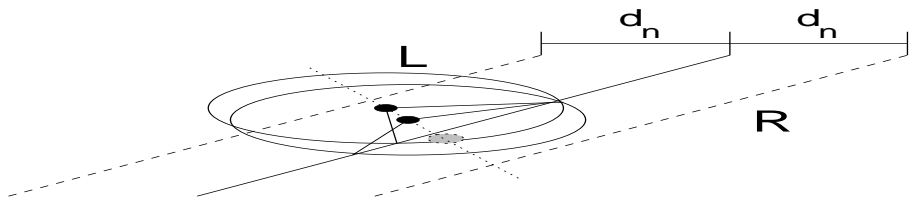,width=12cm,height=2.5cm}}
\vspace{-1mm}
\caption{\small To illustrate how we could end up losing edges: if the
  two black transmitters attempt simultaneously to communicate with the
  gray receiver, a collision will occur, and none of the edges will be
  created.}
\label{fig:manybeams-lost-edges}
\end{figure}

Consider now a fixed transmitter located at some coordinates $(x,y)$.
Any other transmitter located at coordinates $(x',y')\neq(x,y)$ defines
a unique straight line that goes through $(x,y)$ and $(x',y')$.  If
there is a receiver on the other side of the cut along this line, within
reach of both transmitters, then those two edges will be lost---and those
will be the {\em only} lost edges, from among the $\kappa_\vartheta\log(n)$
that each transmitter has.  This situation is illustrated in
Fig.~\ref{fig:manybeams-lost-edges}.

And then we are done.  We have established that in almost all network
realizations, there are $\Theta(nd_n)$ transmitters within each side of
the cut, that each transmitter can reach $\Theta(d_n^2)$
receivers on the other side of the cut, and that integrating out
$\kappa_\vartheta$ we obtain exactly $\Theta(n^2d_n^3)$ edges going
across the cut.  Therefore, the actual number of edges across the cut
is sharply concentrated around its mean, qed.

\subsubsection{Remarks}

\paragraph{MST in a Minimally Connected Network}
Substituting for $d_n = \sqrt{\frac{c\log n}{n}}$ in $\frac 2 3n^2d_n^3$,
we get
\[
  \mbox{\small $\frac 2 3$}n^2\left(\frac{\log n}{\pi n}\right)^\frac{3}{2}
    = \mbox{\small $\frac 2 3$}\sqrt{n} \log^\frac{3}{2}n
    = \Theta\left(\sqrt{n}\log^\frac{3}{2}(n)\right)
\]

Comparing this expression to the ones obtained in
Sections~\ref{sec:omnidirectional} and~\ref{sec:single-directed-beam},
we see that the MST gain due to the use of multiple simultaneous,
arbitrarily narrow beam is, at most, $\Theta\left(\log^2(n)\right)$.

\paragraph{Minimum Connectivity Radius Resuting in MST = $\Theta(n)$}
The minimum $d_n$ resulting in linear MST is obtained by
solving for $d_n$ in $\Theta(n^2 d_n^3) = \Theta(n)$.  Now, for $n$ large
enough, there exist constants $c_1<c_2\in\mathbb{R}$
($c_1>0$ and $c_2<\infty$), such that $c_1 n < \frac 2 3n^2 d_n^3 < c_2 n$,
or equivalently,
$c_1\frac 3 2 n^{-\frac 1 3} < d_n < c_2\frac 3 2 n^{-\frac 1 3}$.
Therefore,
\[ d_n = \Theta(n^{-\frac{1}{3}}). \]
 
\paragraph{Minimum Number of Simultaneous Beams}
In Section~\ref{sec:single-directed-beam}, we said that keeping the
transmission range constant resulted in an impractically large number
of beams that the receiver needed to generate, if linear MST was to
be achieved by increasing the complexity of the signal processing
algorithms.  But if we generate multiple beams, we have just shown
that this minimum radius now is no longer a constant, but instead
tends to zero as $ \Theta(n^{-\frac{1}{3}})$.  However, the situation
is not much better compared to the single beam case, and to see this
again we compute the minimum number of simultaneous beams that a
transmitter would have to generate.

If now we consider the larger $d_n$ satisfying the requirement of
achieving maximum stable throughput linear in network size, then
\[
  \beta = n\cdot\pi d_n^2 = n\Theta(n^{-\frac 2 3}) = \Theta(n^{\frac 1 3}).
\]
Therefore, we see that while $\beta$ is smaller than in the case of the
single beam, we still have an {\em exponential} increase relative
to the number required to maintain minimum connectivity---so again,
we claim that directional antennas are not able to provide an effective
means of overcoming the issue with per-node vanishing throughputs.

\section{Conclusions}
\label{sec:conclusions}

\subsection{Summary of Contributions}

In this paper, we have showed how network flow methods can be used
to determine (to within constants) the maximum stable throughput
achievable in a wireless network.  This was done by formulating MST
as a maximum multicommodity flow problem, for which tight upper and
lower bounds were found.  In the process, the difficult multicommodity
problem was proved equivalent to a simpler single commodity problem,
solvable using standard arguments based on flows and cuts.

As mentioned in the Introduction, this work grows out of our desire
to cast what we deem to be the most useful insights in~\cite{GuptaK:00}
(basically, that the constriction in capacity results from the need
to share constant capacity links by a growing number of nodes), in
a form that makes more intuitive sense to us.\footnote{However, we
certainly would have not been able to obtain our simpler and more
general proofs without the insights provided by cultivating an
appreciation for the line of reasoning employed by Gupta and Kumar
in~\cite{GuptaK:00}.}  And we feel we have accomplished that:
\begin{itemize}
\item By reducing the problem to counting the average number of
  edges that cross a cut and then proving a sharp concentration
  result around this mean, the computational task becomes very
  simple, involving only elementary tools from combinatorics
  and discrete probability.  In~\cite{GuptaK:00}, similar results
  had been obtained based essentially on generalizations of the
  Glivenko-Cantelli theorem (that add uniformity to convergence
  in the law of large numbers), due to Vapnik and
  Chervonenkis~\cite[Ch.\ 2]{Pollard:84}, \cite{VapnikC:71}.
\item In our formulation, it is straightforward to see that
  capacity limitations arise essentially from the geometry of
  the problem: edges have a constant capacity, and only about
  $\sqrt{n}$ of them are available at a minimum cut to transport
  the traffic generated by $n$ sources.
\end{itemize}

%
%

\subsection{Future Work}

In terms of future work, there are a number of interesting
questions opened up by this work.  One deals with the generalization
of these results to nodes distributed on arbitrary manifolds (instead
of the square $[0,1]\times[0,1]\subset\mathbb{R}^2$).  Another
deals with exploring other combinatorial structures (such as
hypergraphs~\cite{Berge:87}), to develop better collision models,
especially in the omnidirectional case.  Of particular interest
to us however is the development of a purely information-theoretic
formulation for the results in this paper, by exploiting the
connections between Shannon information and network flow theory
discovered in~\cite{BarrosS:03e}.

\section*{Acknowledgements}

Will appear in the final version.


\end{document}